\documentclass[aps,pra,amsmath,amssymb,notitlepage,nofootinbib,twocolumn]{revtex4-1}
\usepackage{mathtools,xparse}
\usepackage[utf8]{inputenc}
\usepackage{graphicx,hyperref,upgreek,bbm,mdframed,siunitx}
\usepackage{xcolor}
\hypersetup{    colorlinks,    linkcolor={blue!50!black},    citecolor={blue!50!black},    urlcolor={blue!80!black}}
\usepackage[all]{hypcap}
\usepackage{bbold}
\graphicspath{{img/}}
\newcommand*{\bra}[1]{\ensuremath{\langle #1 \vert}}
\newcommand*{\ket}[1]{\ensuremath{\vert #1 \rangle}}
\newcommand{\braket}[2]{\langle #1 | #2 \rangle}
\newcommand{\ketbra}[2]{| #1 \rangle \langle #2 |}

\newcommand{\mc}[1]{\mathcal{#1}}

\renewcommand{\v}[1]{\textbf{#1}}

\begin{document}

\title{Quantum trajectory entanglement in various unravelings of Markovian dynamics}
\author{Tatiana Vovk}
\email{tatiana.vovk@uibk.ac.at}
\affiliation{Institute for Theoretical Physics, University of Innsbruck, 6020 Austria}
\affiliation{Institute for Quantum Optics and Quantum Information, Austrian Academy of Sciences, 6020 Austria}
\author{Hannes Pichler}
\affiliation{Institute for Theoretical Physics, University of Innsbruck, 6020 Austria}
\affiliation{Institute for Quantum Optics and Quantum Information, Austrian Academy of Sciences, 6020 Austria}
\date{\today}

\begin{abstract}
The cost of classical simulations of quantum many-body dynamics is often determined by the amount of entanglement in the system. In this paper, we study entanglement in stochastic quantum trajectory approaches that solve master equations describing open quantum system dynamics. First, we introduce and compare adaptive trajectory unravelings of master equations. Specifically, building on Ref.~[Phys.~Rev.~Lett.~128,~243601~(2022)], we study several greedy algorithms that generate trajectories with a low average entanglement entropy. Second, we consider various conventional unravelings of a one-dimensional open random Brownian circuit and locate the transition points from area- to volume-law-entangled trajectories. Third, we compare various trajectory unravelings using matrix product states with a direct integration of the master equation using matrix product operators. We provide concrete examples of dynamics, for which the simulation cost of stochastic trajectories is exponentially smaller than the one of matrix product operators.
\end{abstract}
\maketitle

\section{Introduction}
Entanglement plays a central role in modern quantum science as a resource for several quantum information processing tasks~\cite{wootters1998quantum,amico2008entanglement,horodecki2009quantum, calabrese2009entanglement,eisert2010colloquium}. At the same time it poses a substantial obstacle to the efficient simulations of quantum many-body systems on classical computers~\cite{vidalEfficientClassicalSimulation2003,eisert2006general,prosen2007efficiency,schuch2008entropy}.
For instance, simulating quench dynamics of closed many-body systems can lead to the accumulation of extensive entanglement, resulting in prohibitively large cost of classical simulations~\cite{calabrese2007quantum,prosen2007efficiency,schuch2008entropy,polkovnikov2011colloquium,eisert2015quantum,nahum2017quantum}.
The hardness of classically simulating such dynamics has been rigorously established via arguments grounded in complexity theory~\cite{aaronson2005quantum,aaronson2016complexity,bouland2019complexity,napp2022efficient} and has spurred recent interest in developing and implementing protocols aimed at achieving quantum advantage~\cite{bremner2016average,harrow2017quantum,boixo2018characterizing,arute2019quantum,daley2022practical}.

Experimental implementations of quantum many-body dynamics on contemporary quantum hardware are however inherently affected by noise, and thus they are best described in terms of open quantum systems~\cite{de2013purifications,yung2017can,preskillQuantumComputingNISQ2018a,cross2019validating,weimerSimulationMethodsOpen2021}. When the quantum system is open, the complexity of its classical simulation requires reevaluation, since simulating noisy quantum dynamics on a classical computer can potentially be more efficient than its noiseless counterpart. Several classical algorithms for open system dynamics leverage this fact, albeit mostly indirectly. Among them are algorithms based on matrix product or neural network density operators that represent and evolve the full system density operator~\cite{wernerPositiveTensorNetwork2016,whiteQuantumDynamicsThermalizing2018,yoshiokaConstructingNeuralStationary2019,vicentiniVariationalNeuralNetworkAnsatz2019, carisch2023efficient}. Another category encompasses stochastic methods, where the presence of noise is incorporated explicitly as a stochastic element in the algorithm. In particular, in quantum trajectory techniques the system density operator and its dynamics are obtained by statistical averaging over pure-state wave functions~\cite{zoller1987quantum,dum1992monte, dalibard1992wave,gardiner1992wave,gisin1992quantum,tian1992quantum,castin1993wave,van1998quantum,plenio1998quantum}. Tensor network methods such as matrix product states (MPSs) can sometimes provide an efficient means to represent each of these many-body pure-state trajectories~\cite{schollwock2011density,orus2014practical,daley2014quantum,orus2019tensor}.

In Ref.~\cite{vovk2022entanglement} we have introduced a novel way to directly leverage noise in trajectory-based stochastic methods. The central premise relies on the fact that the same system dynamics can be obtained by different unravelings, which give different ensembles of pure-state trajectories. Specifically, Ref.~\cite{vovk2022entanglement} puts forward the idea to adaptively optimize the unraveling choice to minimize the average entanglement, which acts as a proxy of the cost of classically representing trajectories. The physical mechanism underlying this idea is reminiscent of the phenomenon of measurement-induced phase transitions~\cite{li2018quantum,skinner2019measurement,chan2019unitary,li2019measurement,choi2020quantum,zabalo2020critical,jian2020measurement,turkeshi2020measurement,bao2020theory,van2022monitoring,sharma2022measurement,chen2023optimized,cheng2023efficient,hauser2023continuous}. This association stems from the interpretation of the quantum trajectory methods as simulations of the dynamics of (continuously) monitored quantum systems~\cite{holland1996measurement,gardiner2015quantum,fuji2020measurement}.

In this paper, we expand upon the concept introduced in Ref.~\cite{vovk2022entanglement}, analyzing entanglement in various quantum trajectory schemes. We complement our discussion with explicit examples of one-dimensional open quantum dynamics, demonstrating that some trajectory-based methods employing MPSs can yield an exponential reduction in classical computational cost compared to other MPS trajectory-based methods or compared to conventional matrix product operator (MPO) techniques. Finally, we note that our findings are interesting not only from a computational point of view, but also from a fundamental quantum-information-theoretic perspective. This follows the fact that our analysis gives rise to heuristic algorithms for finding upper bounds on mixed-state entanglement measures, such as the entanglement of formation (EoF), a task that holds an independent and intrinsic interest~\cite{audenaert2001variational,guevara2014average,arceci2022entanglement,holevo2022optimization,holevo2023optimization,carisch2023quantifying}.

This paper is structured as follows. In Section~\ref{sec:formalism} we review basic concepts and introduce the formalism this work is built upon. This includes a brief review of the notion of entanglement in pure and mixed states, as well as the discussion on stochastic propagation and quantum trajectories within the formalism of quantum channels. In Section~\ref{sec:res} we show the results obtained using the concepts presented in Section~\ref{sec:formalism}. These include a comparison between various optimization strategies to simulate noisy quantum systems, as well as the results of the numerical experiments on relevant many-body models, with which we explore the practicality of developed ideas. 

\section{Formalism and Methods}
\label{sec:formalism}
In this section we review important basic concepts, introduce the notation and develop the formalism that we employ for the rest of the paper. In Subsection~\ref{susec:entanglement_and_decomposition} we briefly review standard entanglement measures for pure states and efficient tensor network  representations thereof and consider the notions of pure-state ensembles and ensemble-averaged functions of mixed states. In Subsection~\ref{susec:QCQT} we introduce the concept of a quantum channel as well as variations of the quantum trajectory approach as a means to simulate quantum channels. Then in Subsection~\ref{susec:cont_dyn} we consider a special case of continuous Markovian quantum channel, for which in Subsection~\ref{susec:adaptive} we develop the concept of an adaptive entanglement optimization.

\subsection{Bipartite entanglement and pure-state decompositions}
\label{susec:entanglement_and_decomposition}
\subsubsection{Pure states}
\label{susec:pure-state_entanglement}
Consider a pure state $\ket{\psi}$ of a one-dimensional\footnote{Throughout this text we consider one-dimensional systems for simplicity. However, many considerations can be generalized to higher dimensions.} quantum many-body system, consisting of $L$ subsystems with local Hilbert space dimension $d$. Throughout this work we are interested in the entanglement between subsystem $A$ and its complement $B$. Important objects for quantifying this bipartite entanglement are the rank $\chi$ and the eigenvalues $\{\lambda_1, \dots, \lambda_\chi\}$ of the reduced state of either of the partitions. For instance, the von Neumann entanglement entropy is given by~\cite{nielsen2010quantum}:
\begin{align}
    E\left(\ket{\psi}\right) = -\sum_{i=1}^\chi \lambda_i \log_2 \lambda_i.
    \label{eq:vN}
\end{align}
The rank $\chi$, also known as the \textit{bond dimension}, is of central importance for tensor-network-based methods, such as MPS techniques, as it governs the computational cost of representing and processing such states on classical computers~\cite{vidalEfficientClassicalSimulation2003}. The von Neumann entropy of the state~\eqref{eq:vN} is often used as a proxy for this cost. For a comprehensive review on MPSs we refer the reader to Refs.~\cite{schollwock2011density,orus2014practical}.

\subsubsection{Mixed states and pure-state decompositions}
\label{susec:ensembles}
Imagine now that our system is in a mixed state given by the density matrix $\rho$, a positive semi-definite Hermitian operator with a unit trace. This operator can always be decomposed into an ensemble of pure states:
\begin{align}
    \rho =  \sum_{\alpha=1}^k p_\alpha \ket{{\psi}_\alpha}\bra{{\psi}_\alpha}.
    \label{eq:pure-state-decomp}
\end{align}
The decomposition~\eqref{eq:pure-state-decomp} represents $\rho$ as a statistical mixture of pure, normalized states $\ket{\psi_\alpha}$ with probabilities $p_\alpha$. It is also useful to introduce the unnormalized states $\ket{\tilde{\psi}_\alpha}=\sqrt{p_\alpha}\ket{\psi_\alpha}$, such that $\rho =  \sum_{\alpha=1}^k \ket{\tilde{\psi}_\alpha}\bra{\tilde{\psi}_\alpha} 
$, and the notation $\tilde\Psi\equiv(\ket{\tilde{\psi}_1},\ket{\tilde{\psi}_2},\dots,\ket{\tilde{\psi}_k})$ for a compact representation of the pure-state ensemble. Note that the ensemble size $k$ is bounded from below by the rank of the density matrix, $k \geq \mathrm{rank}\left(\rho\right)$.

It is important to notice that the decomposition~\eqref{eq:pure-state-decomp} is not unique~\cite{hughston1993complete}. Two different ensembles of pure states give the same density matrix
\begin{align}
    \rho=\sum_{\alpha=1}^k \ket{\tilde\psi_\alpha}\bra{\tilde{\psi}_\alpha}= \sum_{\alpha=1}^r \ket{\tilde\phi_\alpha}\bra{\tilde{\phi}_\alpha},
    \label{eq:two-pure-state-decomps}
\end{align}
if these two pure-state ensembles are isometrically related:
\begin{align}
\ket{\tilde{\phi}_\alpha}=\sum_{\beta=1}^kT_{\alpha,\beta}\ket{\tilde{\psi}_\beta},
\label{eq:iso}
\end{align} where $T\in\mathcal{T}_k^r$ is an instance of an $r\times k$ isometry (right unitary matrices)~\cite{audenaert2001variational}. We use the short-hand notation $\tilde\Phi=T\tilde\Psi$ to write Eq.~\eqref{eq:iso} more compactly. 

Since the pure-state ensembles $\tilde\Psi$ and $\tilde\Phi$ are equivalent in terms of their density matrix $\rho$, all ensemble-averaged linear functions of the pure-state projectors, \textit{e.g.}, expectation values of operators, coincide. However, the ensemble averages of nonlinear functions are in general different from one ensemble to another. An important example is an ensemble-averaged entanglement entropy (EAEE):
\begin{align}
    \bar{E}[\tilde\Psi]
    = \sum_{\alpha=1}^k  p_\alpha E\left(\ket{{\psi}_\alpha}\right),
    \label{eq:EAEE}
\end{align}
where $E(\ket{{\psi}_\alpha})$ is the von Neumann entanglement entropy~\eqref{eq:vN}. The dependence of the EAEE~\eqref{eq:EAEE} on the pure-state ensemble $\tilde\Psi$ that decomposes $\rho$ in Eq.~\eqref{eq:pure-state-decomp} is the key concept behind the mixed-state entanglement measure known as the entanglement of formation (EoF)~\cite{bennett1996mixed}, which is the result of the minimization of the EAEE over all possible ensembles of all possible sizes $r$:
\begin{align}
    \min_{r} \inf_{T\in\mathcal{T}_k^r} \bar{E}[T\tilde\Psi] \equiv E_f\left(\rho\right).
    \label{eq:EoF}
\end{align}
According to Carath\'{e}odory's theorem~\cite{rockafellar1997convex,uhlmann1998entropy,uhlmann2010roofs}, the optimum~\eqref{eq:EoF} is attained by ensembles of size $r\leq \mathrm{rank}^2\left(\rho\right)$. Computing the EoF for a generic many-body quantum state is known to be NP-hard~\cite{gurvits2003classical,gharibian2008strong}. Nevertheless, finding a pure-state decomposition with low EAEE is practically useful, even if the global minimum can not be found. In particular, in the case when the pure states from~\eqref{eq:pure-state-decomp} are expressed as MPSs, a lower EAEE~\eqref{eq:EAEE} should allow for a more economic state representation.

\subsection{Quantum channels and trajectories}
\label{susec:QCQT}
We describe the dynamics of quantum systems using a quantum channel $\mathcal{E}$~\cite{nielsen2010quantum}, which is a completely positive trace-preserving map that maps an input state of the system $\rho_\mathrm{in}$ to an output state $\rho_\mathrm{out}$:
\begin{align}
\rho_\mathrm{out}=\mathcal{E}(\rho_\mathrm{in})=\sum_{\alpha=1}^k K_{\alpha} \rho_\mathrm{in} K_\alpha^\dag,
\label{eq:QC}
\end{align}
where $K_\alpha$ are the Kraus operators that satisfy $\sum_{\alpha=1}^k K^\dag_\alpha K_\alpha=\mathbb{1}$. It is also useful to introduce the notation $\mathcal{K} = \left(K_1, K_2, \dots, K_k\right)$ as a compact way to denote the Kraus representation of a quantum channel.

It is important to notice that the Kraus representation~\eqref{eq:QC} is not unique. Two different sets of Kraus operators represent the same quantum channel $\mathcal{E}$:
\begin{align}
    \rho_\mathrm{out} = \sum_{\alpha=1}^k K_{\alpha} \rho_\mathrm{in} K_\alpha^\dag = \sum_{\alpha=1}^r R_{\alpha} \rho_\mathrm{in} R_\alpha^\dag,
\end{align}
if they are isometrically related:
\begin{align}
    R_\alpha=\sum_{\beta=1}^k T_{\alpha, \beta} K_\beta,
    \label{eq:iso_Kraus}
\end{align}
with $T\in \mc{T}_k^r$ being an isometric (right unitary) transformation. We use the short-hand notation $\mathcal{R}=T\mathcal{K}$ to write Eq.~\eqref{eq:iso_Kraus} more compactly.
One can notice the analogy between the transformation of the Kraus representations~\eqref{eq:iso_Kraus} and the transformation of the ensemble states~\eqref{eq:iso}. As we will explore below, this transformation between different Kraus representations is directly related to the possibility to construct different stochastic propagation operators for a given quantum channel. 

Throughout this work we are interested in \textit{sequential} quantum channels that are composed of many channels applied in sequence. That is, if the system is initially in a state $\rho_0$, its output state $\rho_n$ after a sequence of $n$ quantum channels can be written as:
\begin{align}
    \rho_n = \mathcal{E}^{(n)}(\rho_{n-1}) = \mathcal{E}^{(n)}(\dots\mathcal{E}^{(i)}(\dots\mathcal{E}^{(1)}(\rho_0))),
    \label{eq:SQC}
\end{align}
where $\mc{E}^{(i)}$ is the $i^\mathrm{th}$ quantum channel given by the Kraus representation $\mathcal{K}^{(i)}=(K_1^{(i)},\dots, K_{k_i}^{(i)})$. This model includes two important examples. First, it can be used to describe discrete quantum circuits, where a single quantum channel $\mc{E}^{(i)}$ represents the the $i^\mathrm{th}$ layer of the circuit. Importantly, the quantum channel description allows to include the case of imperfect quantum gates with uncorrelated noise, an instance of which is considered in Subsubsection~\ref{sususec:MPO_disc}. Second, this model also includes continuous dynamics governed by Markovian master equations, which is described in detail in Subsection~\ref{susec:cont_dyn} and is used in the most parts of Section~\ref{sec:res}. In the remainder of this subsection we consider the connection between sequential quantum channels and quantum trajectories.

\subsubsection{Stochastic propagation and quantum trajectories}
\label{sususec:stoch_traj}
A sequential quantum channel implies a quantum dynamics that can be solved by the quantum trajectory method~\cite{gardiner2015quantum}. In this method, instead of keeping track of the full system state $\rho_n$ from one step to the next, one sequentially samples pure states generated at each step and recovers the solution by statistical averaging. To illustrate this, let us consider a system initialized in a pure state $\rho_0 = \ket{\psi}\bra{\psi}$. The state $\rho_n$ generated by the sequence of quantum channels~\eqref{eq:SQC} can be expressed as
\begin{align}
\rho_n
&=\sum_{\alpha_n,\dots \alpha_1} K^{\left(n\right)}_{\alpha_n} \dots K^{\left(1\right)}_{\alpha_1} \ket{\psi}\bra{\psi}K^{\left(1\right) \dag}_{\alpha_1}\dots K^{\left(n\right) \dag}_{\alpha_n} \label{eq:SQC_state_1}\\
&=\sum_{\alpha_n,\dots, \alpha_1} p_{\alpha_n,\dots ,\alpha_1} \ket{\psi_{\alpha_n,\dots, \alpha_1}} \bra{\psi_{\alpha_n,\dots ,\alpha_1}}, 
\label{eq:SQC_state_2}
\end{align}
where $\alpha_i \in \left\{1, ..., k_i\right\}$ for each $i \in \left\{1, ..., n\right\}$. In the spirit of decomposition~\eqref{eq:pure-state-decomp}, the state~\eqref{eq:SQC_state_2} can be interpreted as a mixture of normalized pure states \begin{align}
\ket{\psi_{\alpha_n,\dots, \alpha_1}}\propto K^{\left(n\right)}_{\alpha_n} \dots K^{\left(1\right)}_{\alpha_1} \ket{\psi}
\label{eq:SQC_trajs}
\end{align}
with probabilities
\begin{align} p_{\alpha_n,\dots \alpha_1} =\bra{\psi}K^{\left(1\right) \dag}_{\alpha_1}\dots K^{\left(n\right) \dag}_{\alpha_n}K^{\left(n\right)}_{\alpha_n} \dots K^{\left(1\right)}_{\alpha_1} \ket{\psi}.
\label{eq:SQC_probs}
\end{align}

The quantum trajectory approach is a method to sample the states~\eqref{eq:SQC_trajs} referred to as quantum trajectories according to the probability distributions~\eqref{eq:SQC_probs}. This is achieved by sequentially sampling pure states at each step of the sequence~\eqref{eq:SQC}. For instance, after application of the first quantum channel from the sequence, $\mc{E}^{(1)}$, the full state of the system is described by
\begin{align}
    \rho_1 =\mc{E}^{(1)}(\rho_0)=\sum_{\alpha_1=1}^{k_1} K_{\alpha_1}^{\left(1\right)} \ket{\psi}\bra{\psi} K_{\alpha_1}^{{\left(1\right)}\dagger},
    \label{eq:single-step-ensemble}
\end{align}
which again, as in~\eqref{eq:pure-state-decomp}, can be viewed as a pure-state decomposition:
\begin{align}
    \rho_1 = \sum_{\alpha_1=1}^{k_1} \ket{\tilde{\psi}_{\alpha_1}} \bra{\tilde{\psi}_{\alpha_1}} = \sum_{\alpha_1=1}^{k_1} p_{\alpha_1} \ket{{\psi}_{\alpha_1}} \bra{{\psi}_{\alpha_1}},
    \label{eq:single-step-ensemble-decomposition}
\end{align}
where $\ket{\tilde{\psi}_{\alpha_1}}=K_{\alpha_1}^{\left(1\right)}\ket{\psi}$ are pure states with corresponding probability weights $p_{\alpha_1}=\braket{\tilde{\psi}_{\alpha_1}}{\tilde{\psi}_{\alpha_1}}$. Using these weights, one can stochastically sample a single pure state from the ensemble $\tilde{\Psi}_1 = \left(\ket{\tilde{\psi}_1}, ..., \ket{\tilde{\psi}_{k_1}}\right)$, thus stochastically propagating the state. That is, with probability $p_{\alpha_1}$ one sets the quantum trajectory after the first step to be $\ket{{\psi}_{\alpha_1}}$.

To continue the stochastic propagation of this trajectory, one executes the same procedure for the next quantum channel, $\mc{E}^{(2)}\left(\ket{{\psi}_{\alpha_1}}\bra{{\psi}_{\alpha_1}}\right)$. Iterating this stochastic algorithm for the whole sequence of channels generates a single quantum trajectory after $n$ steps, $\ket{\psi_{\alpha_n, \dots, \alpha_1}}$, as given in Eq.~\eqref{eq:SQC_trajs}. It is important to observe that the probability to obtain this trajectory is given by~\eqref{eq:SQC_probs}. 

Stochastically generating $M$ independent trajectories according to the algorithm described above, one recovers the state of the system $\rho_n$ from the following unbiased estimator:
\begin{align}
    \rho_n = \lim_{M\rightarrow\infty}\frac{1}{M}\sum_{\ell = 1}^M \ket{\psi_\ell}\bra{\psi_\ell},
    \label{eq:rho_n_sol}
\end{align}
where $\ell = \left(\alpha_n, \dots, \alpha_1\right)$ labels the sampled trajectories. In the case when the initial state is mixed, one can simply consider each pure state from the mixture separately and then take the average.

\subsubsection{Adaptive stochastic propagation}
\label{sususec:adapt_stoch_traj}
In the standard trajectory sampling method described above the Kraus representations are fixed for all of the quantum channels and trajectories. In a more general setting, one is allowed to change Kraus representations from one trajectory to another. This simply follows from the fact that mixtures of different pure-state decompositions of the same density operator also form valid decompositions. We can further generalize this by choosing Kraus representations in a way that depends on the state of current trajectory, that is, in an \textit{adaptive} way~\cite{van1998quantum}. The corresponding adaptive stochastic propagators generate valid quantum trajectories that recover the density matrix via~\eqref{eq:rho_n_sol}, as we discuss in the following.

To demonstrate the validity of adaptive trajectory sampling, let us consider again the first step in the sequence~\eqref{eq:SQC}. Following ~\eqref{eq:iso_Kraus} we choose a transformed Kraus representation $R_{\alpha_1}^{(1)}=\sum_{\beta_1=1}^{k_1} T_{\alpha_1, \beta_1} K_{\beta_1}^{(1)}$ to obtain: 
\begin{align}
    \rho_1 = \sum_{\alpha_1=1}^r R^{(1)}_{\alpha_1} \ket{\psi}\bra{\psi} R^{(1)\dag}_{\alpha_1}.
\end{align}
Since the isometry $T$ is arbitrary, we can choose it in such a way that the corresponding pure-state ensemble $\tilde{\Phi}_1 = (R^{(1)}_{1} \ket{\psi},\dots, R^{(1)}_{r} \ket{\psi})$ has favourable properties. For example, as mentioned in Subsection~\ref{susec:entanglement_and_decomposition}, the states in an ensemble with a lower EAEE~\eqref{eq:EAEE} can be represented more efficiently with MPSs. Importantly, the transformation $T$ can depend on the initial state $\ket{\psi}$. After the choice of $T$ is made, one samples a state $\ket{\tilde{\phi}_{\alpha_1}} = R^{(1)}_{\alpha_1} \ket{\psi}$ with probability $q_{\alpha_1} = \braket{\tilde{\phi}_{\alpha_1}}{\tilde{\phi}_{\alpha_1}}$ and identifies it with the state of the trajectory after the first step.

To continue the propagation of the trajectory $\ket{{\phi}_{\alpha_1}}$, one considers the next quantum channel, $\mc{E}^{(2)}\left(\ket{{\phi}_{\alpha_1}}\bra{{\phi}_{\alpha_1}}\right)$. In the adaptive propagation scheme one chooses the Kraus representation of $\mc{E}^{(2)}$ and the corresponding stochastic propagator based on the state of the \textit{current} trajectory $\ket{{\phi}_{\alpha_1}}$. To see that this is valid and that the resulting trajectories faithfully recover the density matrix~\eqref{eq:rho_n_sol}, let us consider different Kraus representations of $\mc{E}^{(2)}$ for each of the possible trajectories after the first step. That is, we consider Kraus representations that explicitly depend on $\alpha_1$:
\begin{align}R_{\alpha_2|\alpha_1}^{(2)}=\sum_{\beta_2}T_{\alpha_2,\beta_{2}}^{\alpha_1}K_{\beta_2}^{(2)}.
\label{eq:adaptive_trans}
\end{align}
Here $T^{\alpha_1}$ is an isometric transformation that depends on the index $\alpha_1$. It is easy to see that the state after the first two time steps can be equivalently written as:
\begin{align}
\rho_2&=\sum_{\alpha_2\alpha_1}K^{(2)}_{\alpha_2}K^{(1)}_{\alpha_1}\ket{\psi}\bra{\psi}K^{(1)\dag}_{\alpha_1}K^{(2)\dag}_{\alpha_2}\\
&=\sum_{\alpha_2\alpha_1}R^{(2)}_{\alpha_2|\alpha_1}R^{(1)}_{\alpha_1}\ket{\psi}\bra{\psi}R^{(1)\dag}_{\alpha_1}R^{(2)\dag}_{\alpha_2|\alpha_1}\label{eq:ad_sto_pro}.
\end{align}
Eq.~\eqref{eq:ad_sto_pro} indeed shows that the adaptive stochastic propagation scheme described above generates trajectories that faithfully recover the state of the system  from~\eqref{eq:rho_n_sol}. This argument can be easily generalized to prove that an $n$-step adaptive stochastic propagation scheme generates trajectories from a valid decomposition of $\rho_n$ with the proper probability distributions~\cite{vovk2022entanglement}.

\subsection{Continuous dynamics}
\label{susec:cont_dyn}
In this work we are particularly interested in the continuous time dynamics of Markovian quantum processes. If $t$ is the process duration, then the corresponding Markovian quantum channel reads:
\begin{align}
    \rho_t = \mathcal{E}_{t}\left(\rho_0\right) =  e^{\mathcal{L} t} \rho_0,
    \label{eq:MQC}
\end{align}
where $\mathcal{L}$ is the generator of the system dynamics known as the Lindbladian~\cite{lindblad1976generators}, which can be written as\footnote{Here we assume a time-independent Lindbladian for notational simplicity. Generalizations to time-dependent master equations are straightforward. In Section~\ref{sec:res} the time-dependence is indicated explicitly when necessary.}
\begin{align} 
    \mathcal{L} ~\boldsymbol{\cdot}  = -i\left[H, \boldsymbol{\cdot}\right] + \sum_{j=1}^m \gamma_j \left(c_j \boldsymbol{\cdot} c_j^\dagger - \frac{1}{2}\left\{c_j^\dagger c_j, \boldsymbol{\cdot}\right\}\right),
    \label{eq:ME}
\end{align}
where $H$ is a Hamiltonian, $c_j$ are arbitrary operators known as jump (or Lindblad) operators, $\gamma_j \geq 0$ are the associated decoherence rates and $m$ is a positive integer. In order to integrate the dynamical map~\eqref{eq:MQC}, it is often convenient to use the \textit{semigroup} property and discretize the quantum channel into small time steps $dt$~\cite{gorini1976completely}:
\begin{align}
    \rho_t = \mathcal{E}_{t}\left(\rho_0\right) = \underbrace{\mathcal{E}_{dt}(\mathcal{E}_{dt}...(\mathcal{E}_{dt}(}_\text{$n$ times}\rho_0))),
    \label{eq:QC_semigroup}
\end{align}
with $t = n dt$ and $\mathcal{E}_{dt} = \exp\left(\mathcal{L} dt\right)$. Thus, a Markovian quantum channel can be written as a sequential quantum channel~\eqref{eq:SQC} composed of $n$ identical quantum channels $\mathcal{E}^{(i)} = \mathcal{E}_{dt}~\forall i$. For sufficiently small $dt$ one can express $\mc{E}_{dt}$ with a representation consisting of $m+1$ Kraus operators:
\begin{align}
    \left.
        \begin{array}{ll}
            K_0&= \mathbb{1}-i \left(H - \frac{i}{2}\sum_{j=1}^{m} \gamma_{j} c_{j}^\dagger c_{j}\right) dt,~\\
            K_j &= \sqrt{\gamma_j dt} c_j\quad~\mathrm{for} ~j\in\left[1, m\right].
        \end{array}
    \right.
\end{align}
One can check that applying these Kraus operators in~\eqref{eq:QC} with $\alpha\in[0,~m]$ indeed recovers the action of the channel $\mathcal{E}_{dt}$ up to $\mathcal{O}(dt^2)$.

The discretized propagation explained above can be simplified even further by Trotterization~\cite{trotter1959product} of each single-step quantum channel $\mathcal{E}_{dt}$:
\begin{align}
    \mathcal{E}_{dt} = \mathcal{E}_{dt}^\mathrm{coh} \mathcal{E}_{dt}^\mathrm{incoh},
    \label{eq:Trotter}
\end{align}
where the coherent channel $\mathcal{E}_{dt}^\mathrm{coh}$ is unitary and thus has only one Kraus operator $K^\mathrm{coh} = \exp{\left(-i H dt\right)}$, while the incoherent channel can be further split into a sequence of $m$ individual channels $\mathcal{E}_{dt}^\mathrm{incoh} = \mathcal{E}_{dt}^{(1)} ... \mathcal{E}_{dt}^{(j)} ... \mathcal{E}_{dt}^{(m)}$~\cite{daley2014quantum},
where every channel $\mathcal{E}_{dt}^{(j)}$ can be represented with two Kraus operators:
\begin{align}
    \left.
        \begin{array}{ll}
            K^{(j)}_{0}&= \mathbb{1}-\left({\gamma_j dt}/{2}\right) c_j^\dagger c_j,\\
            K^{(j)}_1&= \sqrt{\gamma_j dt} c_j,
        \end{array}
    \right.
    \label{eq:Kraus_local}
\end{align}
that form the Kraus representation $\mathcal{K}^{(j)}_{dt}$. With the decompositions performed above, the Markovian channel \eqref{eq:MQC} is now expressed as a sequential application of $n(m+1)$ simple channels, each of them given either by a unitary channel $\mathcal{E}_{dt}^\mathrm{coh}$ of rank $k = 1$ or by a  individual channel $\mathcal{E}^{(j)}_{dt}$ of rank $k = 2$. Thus, the application of any of these channels to a pure state results in a mixed state whose rank is at most two. This simplicity allows for an optimization of the adaptive stochastic propagation introduced in~\ref{sususec:adapt_stoch_traj}, which we explore in the next subsection. 

\subsection{Adaptive entanglement optimization}
\label{susec:adaptive}
In this section we discuss strategies to \textit{adaptively} choose propagators to find trajectories with a low value of the EAEE~\eqref{eq:EAEE}. As outlined in Subsection~\ref{susec:entanglement_and_decomposition}, such pure-state ensembles are more economic in terms of their MPS representations. Since finding the global EAEE optimum~\eqref{eq:EoF} is NP-hard~\cite{gurvits2003classical,gharibian2008strong}, we devise simpler, heuristic strategies that find ensembles with low (albeit not necessarily minimal) EAEE. Specifically, we exploit the fact that we consider sequential quantum channel to design \textit{greedy} algorithms that break down the global optimization problem into a sequence of simpler problems.

The central idea underlying our greedy optimization algorithms is to propagate pure-state trajectories adaptively by choosing stochastic propagators for each individual channel in such a way that the EAEE is minimised. Let us specify this idea on the example of a continuous Markovian dynamics~\eqref{eq:MQC}, which can be broken down into a sequence of simple channels, as explained in Subsection~\ref{susec:cont_dyn}. For each of these simple channels the pure state of the trajectory, which we denote as $\ket{\psi}$, is stochastically propagated by Kraus operators that are specified by an isometry $T$ (see Subsection~\ref{sususec:adapt_stoch_traj}). We thus choose $T$ such that the EAEE of the trajectory after this propagation step is as small as possible. Therefore, in analogy with Eq.~\eqref{eq:EoF}, we define the following greedy entanglement optimization (GEO) problem\footnote{In what follows we refrain from writing the individual decoherence channel index $j$, whenever there is no potential for confusion.}:
\begin{align}
    \inf_{T\in\mathcal{T}_2^4} \bar{E}[T \tilde{\Psi}_{dt}],
    \label{eq:E_new}
\end{align}
where the reference ensemble $\tilde{\Psi}_{dt} = \mathcal{K}_{dt}\ket{\psi} = (K_0\ket{\psi}, K_1\ket{\psi})$ is generated by the stochastic propagator defined in Eq.~\eqref{eq:Kraus_local} and optimization is performed over the isometries of the class $\mathcal{T}_2^4$, which, according to Carath\'{e}odory's theorem~\cite{rockafellar1997convex,uhlmann1998entropy,uhlmann2010roofs}, is sufficient to attain the minimum. 

Since we work with a Trotterization of a Markovian quantum channel, it is natural to consider the limit of small time steps $dt$. This motivates the definition of the following quantity: 
\begin{align}
    \lim_{dt\rightarrow 0} \inf_{T\in\mathcal{T}_2^4} \frac{ \bar{E}[T \tilde{\Psi}_{dt}] - E\left(\ket{\psi}\right)}{dt} \equiv \dot{\bar{E}}_\mathrm{GEO}.
    \label{eq:E_dot_new}
\end{align} 
Solving the minimization problem~\eqref{eq:E_new} at each step of the propagation sequence thus defines an adaptive trajectory propagation algorithm. Note that each state propagation is optimized independently, with the goal to minimize the entanglement immediately after the propagation step. In this sense the algorithm optimizes locally and in a greedy way. Below we consider several simplifications that are obtained by replacing the cost function~\eqref{eq:E_new} with proxies that are numerically more tractable. 

A way to simplify the optimization~\eqref{eq:E_new} is to consider a limited class of isometries to optimize over. The minimal non-trivial class of isometries is $\mathcal{T}^2_2 \subset \mathcal{T}^4_2$, the instances of which are $2\times2$ matrices $T(\theta, \varphi)$ that can be written as~\cite{hedemann2013hyperspherical}:
\begin{align}
    T(\theta, \varphi) = 
    \begin{pmatrix}
        \cos(\theta) && \sin(\theta) e^{-i\varphi}\\
        \sin(\theta) && - \cos(\theta) e^{-i\varphi}\\
    \end{pmatrix}.
    \label{eq:2x2_T}
\end{align}
With this transformation the optimization reduces to a simple two-parameter optimization over $\theta$ and $\varphi$. Clearly, the result of this optimization gives an upper bound on \eqref{eq:E_dot_new}. In the following we refer to this optimization simply as 2-GEO. The Supplementary Materials (SM)~\ref{SM_sec:T22} and~\ref{SM_sec:T24} contain the details regarding the numerical implementations of the optimization method.

An even simpler algorithm can be obtained by considering the time derivative of the EAEE for a given isometry $T$:
\begin{align}
    \dot{\bar{E}} \left(T\right) \equiv \lim_{dt\rightarrow 0} \frac{\bar{E}[T \tilde{\Psi}_{dt}] - E\left(\ket{\psi}\right)}{dt}.
    \label{eq:EAEE_change_rate}
\end{align}
Minimizing this time derivative with respect to $T$ defines an alternative greedy algorithm to construct trajectories with small average entanglement, which was explored in Ref.~\cite{vovk2022entanglement}. Note that the corresponding greedy entanglement derivative optimization (GEDO) problem
\begin{align}
    \inf_{T\in\mathcal{T}_2^4} \dot{\bar{E}} \left(T\right) \equiv \dot{\bar{E}}_\mathrm{GEDO},
    \label{eq:E_dot_old}
\end{align}
is obtained formally by exchanging the limit and minimization in Eq.~\eqref{eq:E_dot_new}, which implies that the quantity~\eqref{eq:E_dot_old} is an upper bound to the minimum~\eqref{eq:E_dot_new}:
\begin{align}
    \dot{\bar{E}}_\mathrm{GEDO} \geq \dot{\bar{E}}_\mathrm{GEO}.
    \label{eq:old_new_ineq}
\end{align}
Even though GEO~\eqref{eq:E_dot_new} can give trajectories with lower average entanglement, GEDO~\eqref{eq:E_dot_old} has a key feature: $\dot{\bar{E}}_\mathrm{GEDO}$ can be calculated efficiently when the state $\ket{\psi}$ is represented by an MPS. The reason for this is twofold. First, in the GEDO the minimum is attainable with isometries from $\mathcal{T}^2_2 \subset \mathcal{T}^4_2$. Second, the minimization landscape in this case is trivial and can be explored analytically. Specifically, the minimization reduces to choosing between the transform parameters $\theta = 0$ and $\theta = \pi / 4$ and evaluating the optimal phase parameter $\varphi$ (see~\cite{vovk2022entanglement} for details). The evaluation of the analytic expressions of GEDO has a cost of $\mathcal{O}(\chi^3 d)$, where $\chi$ is the bond dimension and $d$ is the local Hilbert space dimension. In contrast, calculating $\dot{\bar{E}}_\mathrm{GEO}$ in general is very costly, since it requires gradient-descent-type numerical optimization, for which we need a normalization sweep at every optimizer iteration. This operation has a cost of $\mathcal{O}(\chi^3 d^3 L \kappa)$, where $L$ is the system length and $\kappa$ is the number of gradient descent iterations (see SM~\ref{SM_sec:T22} for technical details).

\section{Results}
\label{sec:res}
In this section we analyse three distinct aspects of the quantum trajectory algorithms introduced above. First, in Subsection~\ref{susec:minimal}, we apply the adaptive unraveling schemes to simple two-qubit systems. This serves to illustrate differences between optimization schemes as well as their greedy nature. Second, in Subsection~\ref{susec:RBC}, we show that different unraveling schemes can lead to qualitatively different behaviour in the entanglement of many-body systems. For this we consider noisy random Brownian circuits and determine the parameter regimes that correspond to different entanglement scaling laws. We locate the transition between area- and volume-law phases and show that the transition points depend on the unraveling scheme. Third, in Subsection~\ref{susec:MPO}, we compare trajectory-based MPS calculations with standard MPO-based approach and show that the former can be more efficient. For this we construct an explicit example of a noisy many-body models, that can be unraveled using MPSs with a finite bond dimension, while the MPO solution of the same model requires exponential bond dimensions. 

\subsection{Comparison of optimization strategies}
\label{susec:minimal}
\subsubsection{Minimal example: Bell state dephasing}

\begin{figure*}[t]\centering
\includegraphics[width=0.7\textwidth]{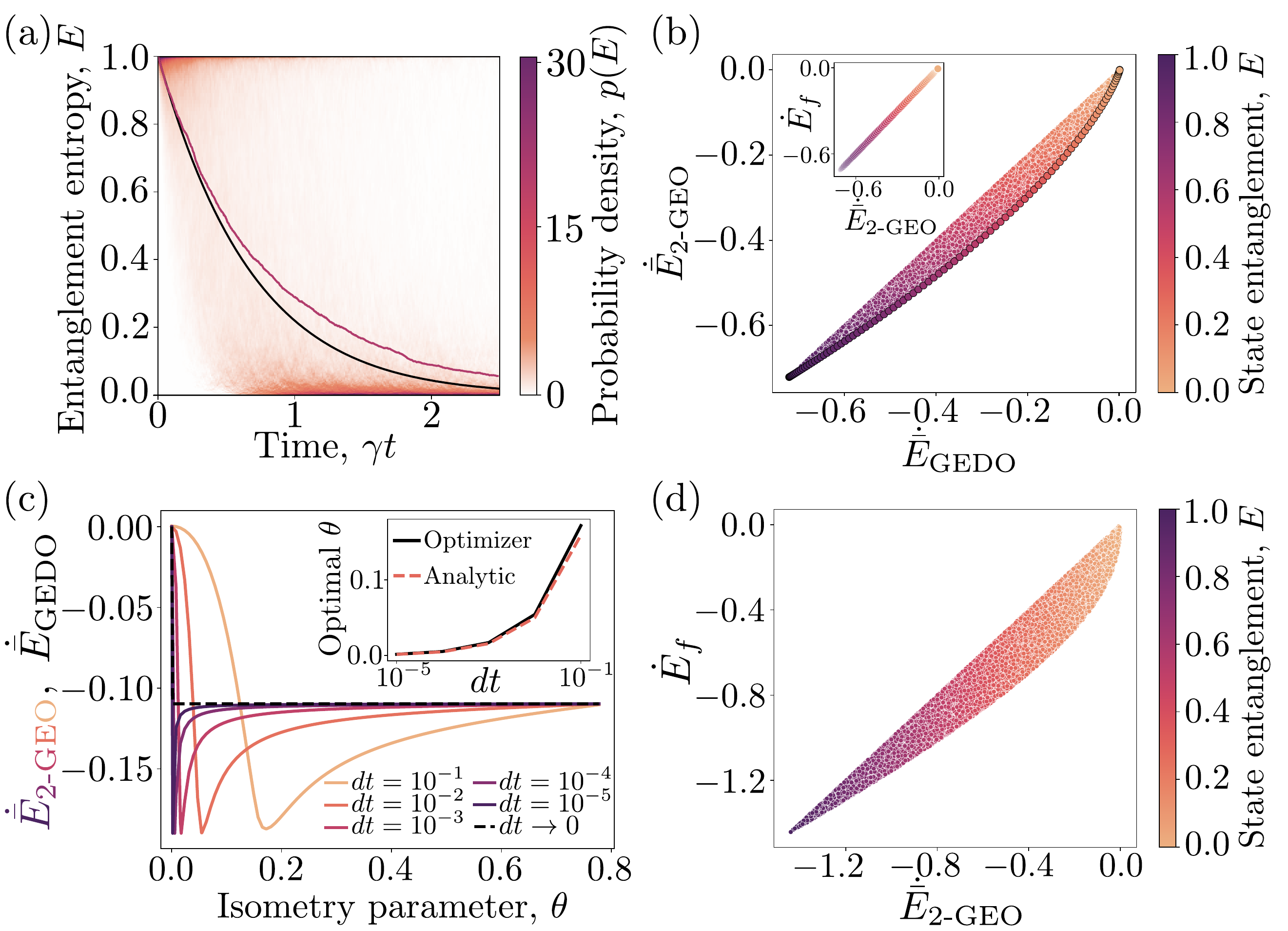}
\setlength\fboxsep{0pt}
\setlength\fboxrule{0.25pt}
\caption{Comparison of optimization strategies. (a) The entanglement probability density distributions over $M=1000$ trajectories for a two-qubit dephasing process with $c_j = \sigma_j^z / 2$. The colour plot corresponds to the distribution of trajectories generated by the GEDO method~\eqref{eq:E_dot_old} with the average indicated by the purple line. The solid black line corresponds to the distribution of the trajectories generated by the 2-GEO method~\eqref{eq:E_dot_new}. The entanglement entropy obtained by the latter method is equal to the EoF~\eqref{eq:EoF} for all the trajectories, therefore its distribution looks like a line.
(b) The EAEE change rates obtained by the 2-GEO vs. GEDO method for a single-channel dephasing of two qubits with $c = \sigma^z / 2$ and $\gamma = 1$. Each dot represents an instance of a Haar random state (total of 20000 instances), the black rims around the dots indicate instances from Eq.~\eqref{eq:x_state} (total of 200 instances). The inset shows the EoF vs. 2-GEO EAEE change rates obtained for the same conditions. The color of the dots corresponds to the initial state entanglement, indicated by the color bars on the right. (c) The EAEE change rates of the 2-GEO (solid colored lines) and GEDO (dashed black line) methods vs. isometry parameter $\theta$ from Eq.~\eqref{eq:2x2_T}. In contrast to the 2-GEO method, the optimization landscape is constant in $\theta$ except the point $\theta = 0$. The inset shows the dependence of the optimal $\theta$ of the 2-GEO method on the value of the time step $dt$, which is in agreement with the analytic formula~\eqref{eq:theta_x_dt}.
(d) The two-qubit EoF vs. the 2-GEO EAEE change rates in the case of two incoherent channels with $c_j = \sigma_j^z / 2$ and $\gamma_j = 1$ ($j=1, 2$). Each dot represents an instance of a Haar random state (total of 20000 instances). The color of the dots corresponds to the initial state entanglement, indicated by a color bar on the right.
}
\label{fig:1}
\end{figure*}

Let us start with a minimal example of two qubits initially in a Bell state, $\ket{\psi} = \ket{\Phi^+}=(\ket{00} + \ket{11}) / \sqrt{2}$. We consider a situation where both qubits dephase with a rate $\gamma$.  That is, their dynamics is described by a Lindblad master equation~\eqref{eq:ME} with $H = 0$ and $c_j = \sigma_j^z / 2$ for $j=1, 2$. The corresponding Kraus operators, as defined in Eq.~\eqref{eq:Kraus_local}, generate trajectories that jump between two Bell states $\ket{\Phi^+}$ and $\ket{\Phi^-} =(\ket{00} - \ket{11}) / \sqrt{2}$, which are both maximally entangled. Hence, the ensemble generated by such an unraveling has the maximal EAEE at all times. In contrast, if we transform the Kraus representation according to Eq.~\eqref{eq:iso_Kraus} with $T(\theta, \varphi)\in\mathcal{T}_2^2$, then it is easy to see that the states in all trajectories are of the form 
\begin{align}
    \ket{\psi(x, \nu)} = \cos(x) \ket{11} + \sin(x) e^{i\nu} \ket{00},
    \label{eq:x_state}
\end{align}
where $x \in\left[0, \pi/2\right]$ and $\nu \in \left[0, 2\pi\right)$. For these states the optimization over $T(\theta, \varphi)$, as defined in  Eq.~\eqref{eq:E_new}, can be explicitly performed. The optimal $\theta$ depends on the state of the trajectory~\eqref{eq:x_state} and is given by 
\begin{align}
    \theta(x, dt) = \arcsin\left[\sec\left(2x\right) \sqrt{\gamma dt}/2\right],
    \label{eq:theta_x_dt}
\end{align}
and the optimal $\varphi = 0$. It is interesting to observe that all the trajectories obtained by this adaptively optimized unraveling have the same entanglement, which matches the EoF during the whole evolution time~\cite{wootters1998entanglement}, as indicated in Fig.~\ref{fig:1}(a) by the black line. We note that this is specific to this minimal example and is not the case in general. In Fig.~\ref{fig:1}(a) we also show the entanglement probability density distribution $p(E)$ for trajectories obtained if we instead minimize the EAEE derivative, given in Eq.~\eqref{eq:E_dot_old}. In this case the entanglement distribution initially spreads and drifts to smaller values before accumulating around zero for longer times. As expected from Eq.~\eqref{eq:old_new_ineq}, in this case the mean entanglement, indicated in Fig.~\ref{fig:1}(a) by the dark purple line, is larger than the EoF. 

\subsubsection{Numerical analysis of single-step optimization strategies}\label{susec:comparison}
In this subsection we consider performance differences and features of the optimization strategies presented above. For this we take several typical states, let them evolve under a decoherence channel for a single time step and compare the resulting EAEE change rates for different channel unravelings. Specifically, we take  Haar-random two-qubit states~\cite{hayden2006aspects} and compare the average entanglement after evolution with 2-GEO and GEDO unravelings of a single decoherence channel. In Fig.~\ref{fig:1}(b) we make such a comparison for the jump operator $c = \sigma^z / 2$. We can see that the 2-GEO method that directly optimizes the entanglement gives lower EAEE values than the GEDO approach that optimizes the entanglement derivative, which is consistent with the inequality~\eqref{eq:old_new_ineq}. We note that the EAEE change rate is larger for the states with larger initial entanglement. Interestingly, the states taken from~\eqref{eq:x_state} highlighted by black rims in Fig.~\ref{fig:1}(b) give the largest difference for the EAEE change rate. In addition, as shown in the inset of Fig.~\ref{fig:1}(b), we also observe that, for the case of two qubits with a single decoherence channel, the 2-GEO method~\eqref{eq:E_new} is sufficient to reach the EoF~\cite{wootters1998entanglement}, which implies that in this case going from GEO to 2-GEO is not a limitation. In Section~\ref{SM_sec:10qubit} of the SM we provide the same analysis for states of more qubits.

To gain more insights into the difference between the two optimization methods, it is useful to consider the optimization landscapes in both cases. For this it is important to note that the optimization landscape of the 2-GEO method~\eqref{eq:E_dot_new} depends on the time step $dt$. In contrast, the GEDO approach~\eqref{eq:E_dot_old} does not depend on $dt$ due to the fact that the limit of $dt\rightarrow 0$ is taken \textit{before} the minimization. To illustrate this, in Fig.~\ref{fig:1}(c) we take a particular instance of the state~\eqref{eq:x_state} with $\cos(x) = 0.98$ as an initial state and consider the EAEE optimization landscapes\footnote{We consider specific cuts of optimization landscapes with a fixed phase parameter at $\varphi = 0$, since in the considered case this contains the global minimum.} for different values of the time step $dt$. Notably, the minimal EAEE values differ for the 2-GEO and GEDO methods. In particular, the minimal value of $\dot{\bar{E}}_\text{2-GEO}$ is below $\dot{\bar{E}}_\mathrm{GEDO}$. This highlights the fact that the the optimization and the limit $dt\rightarrow 0$ do not commute. Additionally, we observe that the GEDO optimization landscape is trivial, as the optimization landscape is constant for every value of $\theta$ except when $\theta = 0$, and the minimum in this case is obtained by comparing the cases of $\theta = 0$ and $\theta = \pi / 4$ (see also Ref.~\cite{vovk2022entanglement}).

\subsubsection{Two-channel propagation analysis}
An important aspect of the adaptive propagation strategies discussed in Subsection~\ref{susec:adaptive} is their greedy nature, \textit{i.e.}, the fact that the optimization is done separately for each decoherence channel~\eqref{eq:Kraus_local} in the (Trotter) sequence. This becomes crucial in the case of multiple decoherence channels, where optimizing each of them separately does not necessarily produce the ensemble with minimal EAEE. We illustrate this again with an example of two independently dephasing qubits. For this setup we compare the entanglement obtained from sequentially optimized unraveling to the global optimum, which is given by the EoF~\cite{wootters1998entanglement}. Specifically, we evolve the two qubits for a single time step $dt$, which consists of a sequential application of a dephasing channels to each of the qubits. That is, we first apply a dephasing channel to the first qubit and then propagate the resulting state with a dephasing channel that acts on the second qubit. Each propagation is optimized independently according to the 2-GEO method. From the resulting states we extract the change rate of the EAEE, which we plot against the entanglement of formation of the mixed quantum state in Fig.~\ref{fig:1}(d). As expected, the sequential optimization of the two channels does not give the minimal configuration.

\subsection{Entanglement in noisy random Brownian circuits}
\label{susec:RBC}
So far we only considered two-qubit systems and evolution for a single time step. For larger system sizes and longer evolution times more interesting phenomena can occur. The competition between unitary dynamics and incoherent processes can give rise to distinct phases, characterized by different entanglement properties of the trajectories.
If the unitary dynamics dominates in a given unraveling scheme, the entanglement of the trajectories can grow extensively with the system size (volume-law phase). In contrast, when the stochastic processes dominate, the entanglement can be bounded by the size of the subsystem boundary (area-law phase)~\cite{eisert2006general,bao2020theory}. This transition, typically referred to as measurement-induced phase transition, has been observed in several models in recent studies~\cite{li2018quantum,skinner2019measurement,chan2019unitary,li2019measurement,choi2020quantum,zabalo2020critical,jian2020measurement,turkeshi2020measurement,bao2020theory,van2022monitoring,sharma2022measurement,chen2023optimized,cheng2023efficient}. Importantly, as shown in Ref.~\cite{vovk2022entanglement}, the transition point can change for different unraveling schemes. Due to the connection between the entanglement and computational cost for classical simulations with tensor networks, different trajectory propagation approaches can thus differ significantly in terms of their computational cost for solving the same master equation.

\begin{figure*}[t]\centering
\setlength\fboxsep{0pt}
\setlength\fboxrule{0.25pt}
\includegraphics[width=0.9\linewidth]{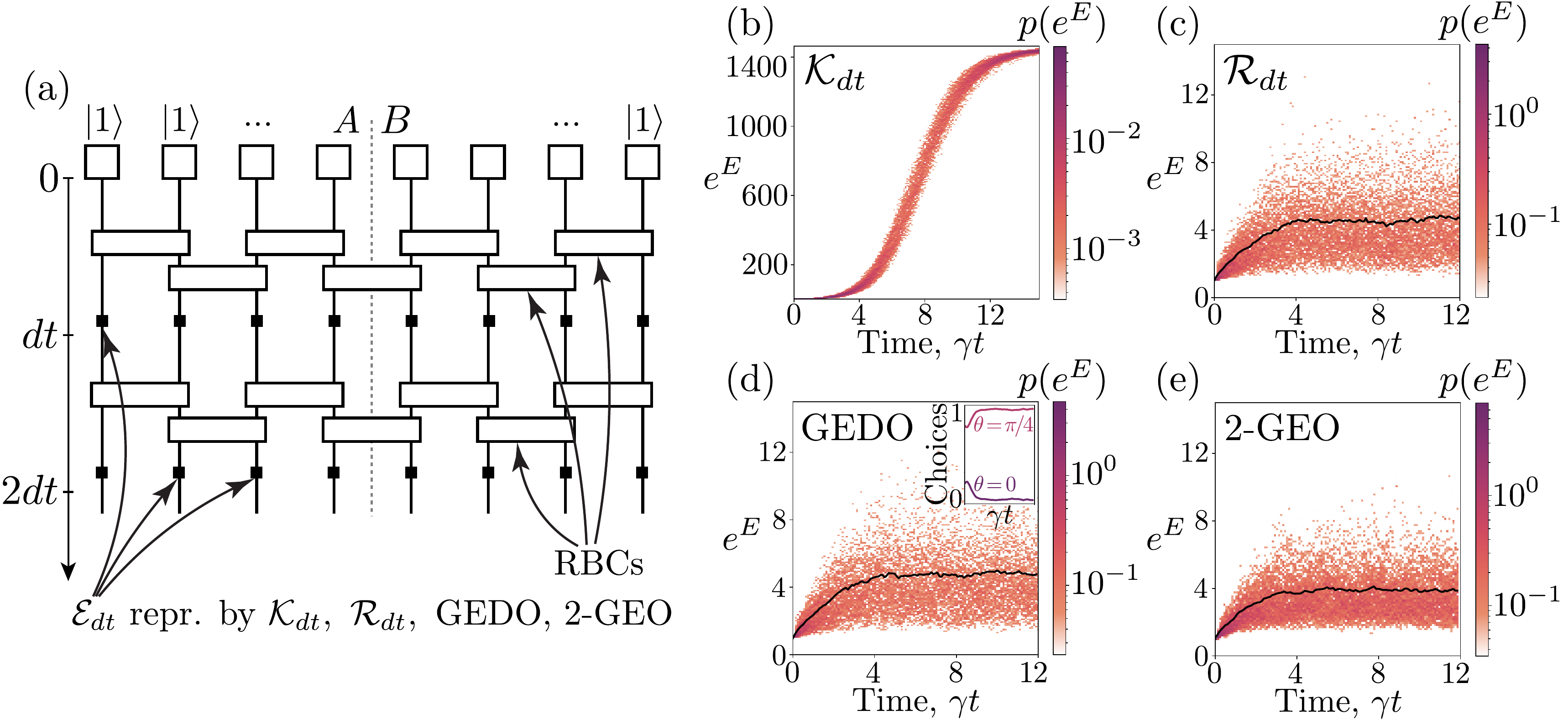}
\caption{Entanglement distributions in various unravelings of the continuous RBC model with dephasing. (a) The initial state $\ket{\psi} = \ket{11 \dots 1}$ is propagated by an open continuous RBC consisting of alternating Trotter layers corresponding to coherent and incoherent part of the master equation. For each incoherence channel $\mathcal{E}_{dt}$ (the channel index $j$ is omitted) we consider the ``quantum jump'' unraveling that uses the standard Kraus representation $\mathcal{K}_{dt}$, ``quantum state diffusion'' unraveling that uses the transformed Kraus representation $\mathcal{R}_{dt}$ (see text), the GEDO method~\eqref{eq:E_dot_old} and the 2-GEO method~\eqref{eq:E_dot_new}. Panels (b) -- (e) show the corresponding entanglement probability density distributions of the entanglement exponent $p\left(e^E\right)$ in the trajectories modelling the RBC dephasing process. For panels (c), (d) and (e) we additionally plot the average of the entanglement exponent on top of the distributions (black lines). The inset in (d) illustrates the average proportion of choices between cases $\theta = 0$ and $\theta = \pi/4$ done by the GEDO algorithm (see text). The number of qubits in the system is $L=16$, the jump operators are $c_j = \sigma_j^z / 2$ and the decoherence rate relative to the variance of the Gaussian variable in the RBC~\eqref{eq:RBC} is $\gamma = 10$ ($\gamma_j = \gamma$). The number of trajectories with various RBC realizations is $M = 200$ for each panel, the bond dimension is $\chi = 256$, which is exact for $L = 16$. We emphasize a different vertical scale in the panel (b) compared to the panels (c), (d) and (e).
}
\label{fig:2}
\end{figure*}

To see this, let us consider the model studied in Ref.~\cite{vovk2022entanglement}, \textit{i.e.}, the noisy Random Brownian circuit (RBC) that is schematically depicted in Fig.~\ref{fig:2}(a). It describes a one-dimensional chain of spin-$1/2$ systems, whose coherent evolution is given by the Hamiltonian:
\begin{align}
   H(t) = \sum_{j=1}^{L-1}\sum_{o, p=0}^3 g^{o, p}_j(t) ~\sigma_j^o \otimes \sigma_{j+1}^p,
   \label{eq:RBC}
\end{align}
where $\sigma_j^o \in \left\{\mathbb{1}_j, \sigma_j^x, \sigma_j^y, \sigma_j^z\right\}$ are the Pauli matrices acting on qubit $j$. The parameters $g^{o, p}_j$ are Gaussian stochastic variables with $\left<\left<g^{o, p}_j\right>\right> = 0$ and $\left<\left<g^{o, p}_j g^{o', p'}_{j'}\right>\right> = \alpha \delta_{j, j'} \delta_{o, o'} \delta_{p, p'}$, where $\left<\left<...\right>\right>$ denotes the average over Hamiltonian realizations and $\alpha$ is the variance of the Gaussian variable\footnote{Without the loss of generality, in what follows we set $\alpha = 1$.}. The incoherent part of the evolution is described by  jump operators $c_j = \sigma_j^z / 2$ with $\gamma_j = \gamma$.

In panels (b -- e) of Fig.~\ref{fig:2} we show the histograms of the entanglement in the various trajectories that solve the dynamics of the noisy RBC. Specifically, we plot the exponential $e^E$ of the entanglement entropy~\eqref{eq:vN} of every trajectory, since this quantity serves as a proxy for the bond dimension $\chi$ required to represent the trajectory in the MPS form~\cite{vidalEfficientClassicalSimulation2003}. In panel (b) of Fig.~\ref{fig:2} we depict the histogram of trajectories generated by the Kraus representation $\mathcal{K}_{dt}$ defined in Eq.~\eqref{eq:Kraus_local}, which corresponds to the standard quantum trajectory method known as the ``quantum jump'' approach~\cite{zoller1987quantum,dum1992monte,dalibard1992wave,gardiner1992wave,castin1993wave,plenio1998quantum}. In this case, both of the propagation operators from Eq.~\eqref{eq:Kraus_local} are local unitary operations, and thus the incoherent part of the evolution does not remove any entanglement, despite the large decoherence rate ($\gamma = 10$). The dynamics of the entanglement is thus determined by the coherent part, which leads generically to linear entanglement growth. This growth continues  until a saturation value proportional to the system size is reached. In Fig.~\ref{fig:2}(c) we plot the histogram generated by the Kraus representation $\mathcal{R}_{dt} = T(\theta, \varphi) \mathcal{K}_{dt}$, where $T(\theta, \varphi)$ defined in Eq.~\eqref{eq:2x2_T} has fixed parameters $\theta = \pi/4$ and $\varphi = 0$. This unraveling is known as the ``quantum state diffusion''~\cite{gisin1992quantum,tian1992quantum,van1998quantum}. In contrast to the ``quantum jump'' method, the ``quantum state diffusion'' approach results in entanglement distributions that saturate at much smaller values. Importantly, these values here are not determined by the system size, but by the decoherence rate (see also Ref.~\cite{vovk2022entanglement}). In Fig.~\ref{fig:2}(d) we depict the histogram obtained by the GEDO method, which minimizes the EAEE change rate~\eqref{eq:E_dot_old} by switching between transformed Kraus representations with $\theta = 0$ and $\theta = \pi / 4$ (with free phase parameter $\varphi$) and selecting the best one at every propagation step. Comparing panels (c) and (d) we observe a similarity in the trajectory distributions. This is due to the fact that the GEDO approach in this case almost exclusively selects $\theta = \pi / 4$ with $\varphi = 0$, which corresponds to the ``quantum state diffusion'' unraveling. This can be also seen from the inset in Fig.~\ref{fig:2}(d), where we show the average proportion of $\theta$ parameter choices made by the GEDO algorithm\footnote{As in the example in Fig.~\ref{fig:1}(c), in this case the GEDO minimum is found always at $\varphi = 0$, therefore this parameter is not shown.}. At last, in Fig.~\ref{fig:2}(e) we plot the histogram obtained by the 2-GEO method~\eqref{eq:E_new}, which minimizes the EAEE by adaptively choosing the transform $T(\theta, \varphi)$. In this case the parameters $\theta$ and $\varphi$ are determined numerically and the optimization cannot be reduced to switching between two cases as it happens in the GEDO method. Among the four unraveling schemes shown in Fig.~\ref{fig:2}, the 2-GEO one results in the trajectories with the smallest average entanglement and also with the smallest weight on trajectories with higher entanglement.

\begin{figure*}[t]\centering\includegraphics[width=1.\textwidth]{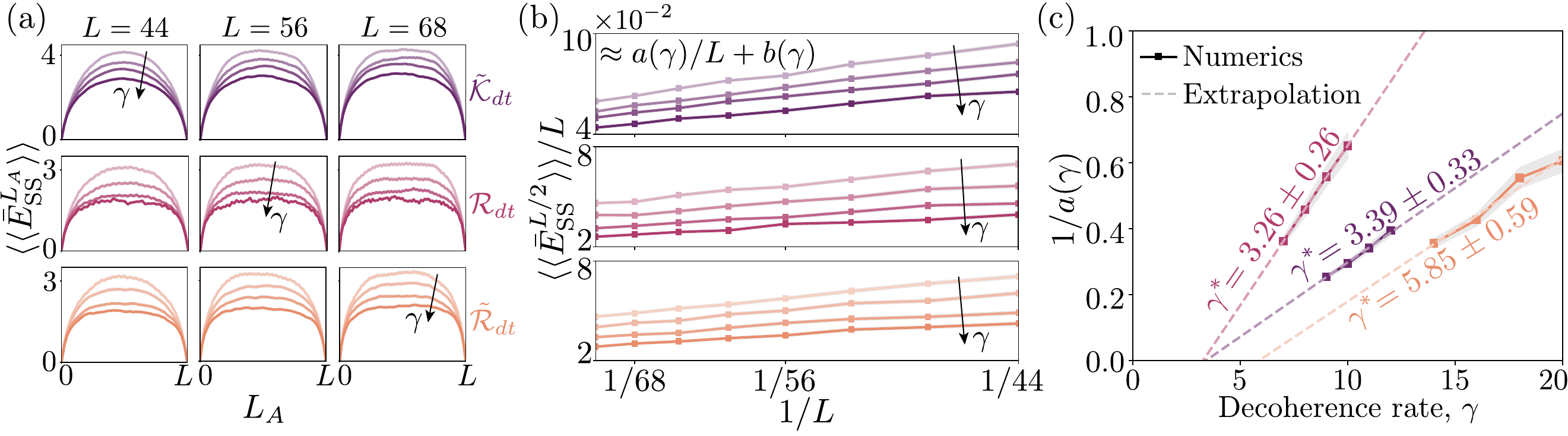}
\setlength\fboxsep{0pt}
\setlength\fboxrule{0.25pt}
\caption{Determination of the critical decoherence rates $\gamma^*$ in the continuous RBC model with dephasing for various unravelings. (a) The dependence of the distributions of the steady-state EAEE $\langle\langle\bar{E}_\mathrm{SS}^{L_A}\rangle\rangle$ over the bipartite cut location $L_A$ on the decoherence rate $\gamma$ for various system sizes $L$ and decoherence channel representations $\tilde{\mathcal{K}}_{dt}$, $\mathcal{R}_{dt}$ and $\tilde{\mathcal{R}}_{dt}$ (see text for more details). The arrows indicate the increasing values of $\gamma$ different for each of the Kraus representations: $\gamma = \left\{9,~10,~11,~12\right\}$ for $\tilde{\mathcal{K}}_{dt}$, $\gamma = \left\{7,~8,~9,~10\right\}$ for $\mathcal{R}_{dt}$ and $\gamma = \left\{14,~16,~18,~20\right\}$ for $\tilde{\mathcal{R}}_{dt}$. (b) The dependence of the middle-cut steady-state EAEE divided by the system size $\langle\langle\bar{E}_\mathrm{SS}^{L/2}\rangle\rangle / L$ on the inverse system size $1/L$. According to the ansatz~\eqref{eq:area-volume}, the slope of the plot corresponds to the area-law coefficient $a(\gamma)$. The arrows indicate the increasing values of $\gamma$. (c) The dependence of the inverse values of the area-law coefficients $1/a(\gamma)$ on the decoherence rate $\gamma$. The data is compatible with linear behaviour, therefore we use the linear extrapolation to determine the critical decoherence rates $\gamma^*$ written in corresponding colors in the plot. The grey shadings indicate the standard errors of the numerical lines, while the dashed lines illustrate the extrapolations. The system sizes used to generate the data: $L  = \left\{44,~48,~52,~56,~60,~64,~68,~72\right\}$, the number of trajectories with various RBC realizations is $M = 200$ and the bond dimension is $\chi = 300$. The bond dimension convergence is checked against instances with $\chi = 512$.
}
\label{fig:3}
\end{figure*}

Importantly, the entanglement distributions do not only differ quantitatively across unraveling schemes, but they also change qualitatively at long times~\cite{vovk2022entanglement}. In particular, for the unraveling in Fig.~\ref{fig:2}(b) the entanglement distribution concentrates around a saturation value that is proportional to the system size. In contrast for unravelings in panels (c -- e) of Fig.~\ref{fig:2}, the saturation value is independent on the system size, indicating that the system is in the area-law phase. This area-law phase is expected to persist for sufficiently large decoherence rate $\gamma$. At the critical value $\gamma = \gamma^\ast$, one expects a so-called measurement-induced phase transition from area- to volume-law entanglement. Hence, interesting quantities to determine are the critical decoherence rates $\gamma^*$ for various unraveling schemes.

In the following we focus on determining the critical points in unravelings congenial to some of the ones presented in Fig.~\ref{fig:2}. The ``quantum jump'' unraveling shown in Fig.~\ref{fig:2}(b) demonstrates a volume-law scaling and hence does not possess a critical point. We therefore consider a variation of ``quantum jump'' unraveling with the following Kraus representation:
\begin{align}
    \left.
        \begin{array}{ll}
            \tilde{K}^{(j)}_{0}&= \mathbb{1}-\left({\gamma dt}/{2}\right) \ket{1}_j \bra{1},\\
            \tilde{K}^{(j)}_1&= \sqrt{\gamma dt} \ket{1}_j \bra{1},
        \end{array}
    \right.
\end{align}
to which we refer to as $\tilde{\mathcal{K}}_{dt}$ for brevity\footnote{This representation can be obtained by applying the isometric transformation~\eqref{eq:2x2_T} with parameters $\theta\left(dt\right) = \arcsin\left(\sqrt{\gamma dt}/2\right)$ and $\varphi = 0$ to the original representation. Note that in this case the transformation is time-step-dependent.}. We also study ``quantum-state-diffusion'' unravelings:
\begin{align}
    \left.
        \begin{array}{ll}
            R^{(j)}_{\pm}&= \frac{1}{\sqrt{2}}\left[\left(1-{\gamma dt}/{8}\right) \mathbb{1}_j \pm e^{-i\varphi} \sqrt{\gamma dt}\sigma_j^z / 2\right],
        \end{array}
    \right.
\end{align}
of which we take two instances with $\varphi = 0$ and $\varphi = \pi / 4$ and consider how changing the phase parameter affects the critical point. We refer to these representations as $\mathcal{R}_{dt}$ and $\tilde{\mathcal{R}}_{dt}$, respectively.

In what follows we do not consider adaptive unravelings shown in Fig.~\ref{fig:2}(d) and (e). The reason for this is twofold. First, as already mentioned, in the case of the open RBC the GEDO approach chooses mostly the ``quantum state diffusion'' unraveling. We therefore expect this method to have similar critical decoherence rate. Second, as mentioned in the end of Subsection~\ref{susec:adaptive}, our implementation of the 2-GEO unraveling is too computationally demanding for large system sizes that are required to establish the critical point (see SM~\ref{SM_sec:T22} for details).

\begin{figure*}[t]\centering
\includegraphics[width=0.8\textwidth]{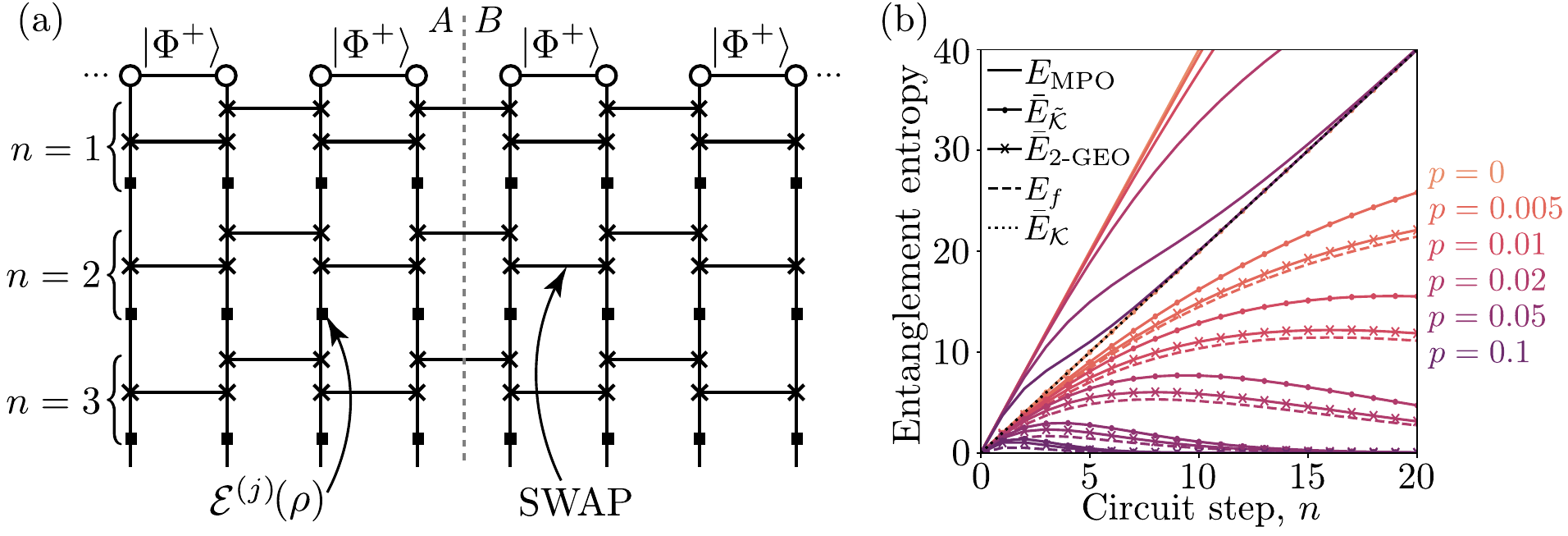}
\setlength\fboxsep{0pt}
\setlength\fboxrule{0.25pt}
\caption{Discrete noisy quantum circuit and entanglement dynamics therein for different simulation methods. (a) Discrete quantum SWAP circuit with dephasing applied to infinite chain of Bell pairs~\eqref{eq:state_discrete}. Each discrete circuit step $n$ consists of nearest-neighbour SWAP gates applied alternatively to odd and even neighbouring qubit pairs and individual decoherence channels $\mathcal{E}^{(j)}$ defined in Eq.~\eqref{eq:QC_discrete}. The system's bipartition $AB$ is shown by the grey dashed line. (b) The (operator) entanglement dynamics during the discrete noisy circuit evolution with various decoherence rates~$p$. The lines' colors that change from bright orange to dark purple correspond to different values of the decoherence rate, $p = 0,~0.005,~0.01,~0.02,~0.05$, and $0.1$. Colored solid lines indicate the operator entanglement~\eqref{eq:OE} and the black dotted line shows the EAEE dynamics for the standard Kraus representation $\mathcal{K}^{(j)}$ from Eq.~\eqref{eq:Kraus_discrete} that is independent on $p$. Both of these cases exhibit an unbounded entanglement growth. Colored solid lines with circles indicate the EAEE~\eqref{eq:EAEE_analytic} of the trajectories generated with the transformed Kraus representation $\mathcal{R}^{(j)}$, while colored solid lines with crosses correspond to EAEE of the 2-GEO method (total of $M = 50$ trajectories). The colored dashed lines illustrate the EoF minimum~\eqref{eq:EoF}.
}
\label{fig:4}
\end{figure*}

To determine the critical decoherence rates $\gamma^*$ for different unravelings, we use a finite-size scaling argument. We consider the entanglement entropy averaged over trajectories and RBC realizations, $\langle\langle\bar{E}^{L_A}_\mathrm{SS}\rangle\rangle$, where $L_A$ is the size of the subsystem. The subscript SS indicates that the value is evaluated at late times, when the entanglement distribution becomes stationary. When $L_A = L/2$, the above quantity is constant in the area-law phase and grows linearly with the system size $L$ in the case of the volume-law scaling. Thus one can make the following ansatz: 
\begin{align}
    \langle\langle\bar{E}^{L/2}_\mathrm{SS}\rangle\rangle  = a\left(\gamma\right) + b\left(\gamma\right) L,  
    \label{eq:area-volume}
\end{align}
where $a\left(\gamma\right)$ is the EAEE saturation value in the area-law phase and $b\left(\gamma\right)$ is the slope of the volume-law scaling. The area-law phase is thus identified when $b\left(\gamma\right)=0$ and $a\left(\gamma\right)$ is finite. When $\gamma$ approaches the critical point $\gamma^*$ from above (\textit{i.e.}, from the area-law phase), we expect $a\left(\gamma\right)$ to diverge, while $b\left(\gamma\right)$ should remain zero. In Fig.~\ref{fig:3} we use this to locate the critical point $\gamma^*$. Specifically, we start with the system initially in the state $\ket{\psi} = \ket{11 \dots 1}$, which we evolve with the random circuit~\eqref{eq:RBC} with dephasing channels as depicted in Fig.~\ref{fig:2}(a). The dephasing channels in the circuit are expressed by the mentioned representations $\tilde{\mathcal{K}}_{dt}$, $\mathcal{R}_{dt}$ and $\tilde{\mathcal{R}}_{dt}$, which result in different trajectory unravelings. 

In Fig.~\ref{fig:3}(a) we plot the EAEE in the system after sufficiently long evolution time, $\langle\langle\bar{E}^{L_A}_\mathrm{SS}\rangle\rangle$, for various system sizes, unraveling schemes and decoherence rates. Using this data, in Fig.~\ref{fig:3}(b) we plot the middle-cut steady-state EAEE against the inverse of the system size, from which, using the ansatz~\eqref{eq:area-volume}, we can extract the area-law coefficients $a\left(\gamma\right)$ and confirm that the volume-law coefficient $b\left(\gamma\right) = 0$ within the fitting error. In Fig.~\ref{fig:3}(c) we plot the inverse of area-law coefficients against the decoherence rate and extrapolate this data to get the critical rates $\gamma^*$.

From the results obtained in Fig.~\ref{fig:3}(c) we observe that the critical points indeed depend on the particular unraveling of the decoherence channels. Specifically, unravelings with Kraus representations $\tilde{\mathcal{K}}_{dt}$ and $\mathcal{R}_{dt}$ give lower critical values compared to the unraveling with $\tilde{\mathcal{R}}_{dt}$. Another observation stems from comparing the ``quantum state diffusion'' unravelings $\mathcal{R}_{dt}$ and $\tilde{\mathcal{R}}_{dt}$, where different values of the phase parameter $\varphi$ significantly affect the location of the critical point. Specifically, within the standard error the obtained critical decoherence rates are related to each other as $\gamma^* |_{\mathcal{R}_{dt}}  = \cos^2\left(\pi/4\right) \gamma^* |_{\tilde{\mathcal{R}}_{dt}}$, as predicted in Ref.~\cite{vovk2022entanglement}.

\subsection{Comparison with other approaches to solve the many-body dynamics}
\label{susec:MPO}
We study entanglement properties in various quantum trajectory unravelings, because this is crucial for determining the computational cost of solving the master equation~\eqref{eq:ME} using quantum trajectories represented with MPSs (which we refer to as QT MPS in the following). Given this motivation, it is important to compare this approach with other existing methods to solve master equations. Specifically, in this subsection we compare the QT MPS method with an MPO approach~\cite{bonnes2014superoperators,wellnitz2022rise,preisser2023comparing}. The latter consists in vectorizing the density matrix and corresponding equations of motion~\eqref{eq:ME} and representing them as tensors. In general, it is unclear which of the two methods, QT MPS or MPO, is computationally less expensive. The computational cost of a simulation based on MPO scales as $\mathcal{O}\left(\chi^3 d^6 L n\right)$, where $\chi$ is the bond dimension, $d$ is the local Hilbert space dimension, $L$ is the system length and $n$ is the total number of (continuous) time steps. In turn, for the QT MPS method (with a fixed statistical accuracy) the computational cost scales as $\mathcal{O}\left(\chi^3 d^3 L n\right)$. Note that the bond dimension $\chi$ has different physical meaning for the MPS and MPO. As discussed above, in the MPS approach the bond dimension $\chi$ is related to the entanglement in individual trajectories. On the other hand, for the MPO the bond dimension represents both quantum and classical correlations in the density matrix. To illustrate this difference we consider a simple model of a system that features unbounded growth of classical correlations during its evolution, while the amount of entanglement in individual trajectories can remain finite. 

\subsubsection{Discrete case}
\label{sususec:MPO_disc}
Let us consider an infinite one-dimensional chain of qubits being initially a product of nearest-neighbour Bell pairs:
\begin{align}
    \ket{\psi} = \dots \otimes\ket{\Phi^+}\otimes\ket{\Phi^+}\otimes\ket{\Phi^+}\otimes\ket{\Phi^+}\otimes \dots.
    \label{eq:state_discrete}
\end{align}
We consider time evolution of this initial state under the discrete circuit depicted in Fig.~\ref{fig:4}(a), which consists of alternating layers of nearest-neighbor SWAP gates~\cite{nielsen2010quantum} and individual local dephasing channels of the form:
\begin{align}
    \mathcal{E}^{(j)}\left(\rho\right) = \left(1-p\right)\rho + p \sigma_j^z \rho \sigma_j^z,
    \label{eq:QC_discrete}
\end{align}
where $p$ is the decoherence strength. 

To understand the dynamics of this model, first note that the coherent part of the evolution (induced by the SWAP gates) simply leads to a ballistic separation of the the initially correlated nearest-neighbor pairs across the chain, with the distance between each partner of a pair growing as $4n$. The incoherent part does not affect this ballistic dynamics, but instead leads to a dephasing of each pair, which transforms entanglement in each pair into classical correlations. The dynamics thus simply reduces to that of independent pairs, and measures of entanglement or correlations across a given bipartition (and the associated computational costs of MPS or MPS representations) can thus be evaluated analytically. 

As a proxy of the computational cost of the MPO method we consider its operator entanglement entropy~\cite{wellnitz2022rise}:
\begin{align}
    &E_\mathrm{MPO} = - \sum_i \mu^2_i \log_2 \mu^2_i,~\text{with}~\mu_i = \frac{\tilde{\mu}_i}{\sqrt{\sum_{i'} \tilde{\mu}_{i'}^2}},
    \label{eq:OE}
\end{align}
where $\tilde{\mu}_i$ ($\mu_i$) are the (normalized) singular values of the MPO's middle cut. In the above setting the corresponding operator entanglement grows with the circuit depth $n$, and has two characteristic regimes. Initially, when $pn \ll 1$, the dephasing processes do not destroy the quantum entanglement in Bell pairs, hence the operator entanglement initially scales as $4n$. For later evolution times, when $pn \gg  1$, only the classical correlations remain, which is reflected in the operator entanglement scaling as $2n$. More generally, the operator entanglement is given by:
\begin{align}
    E_\mathrm{MPO}(p, n) &= 2n + \frac{2n}{1 + \left(1-p\right)^{8n}} \bigg\{\log_2\left[1+\left(1-p\right)^{8n}\right] \nonumber \\
    &+ \left(1-p\right)^{8n}\log_2\left[1+\left(1 - p\right)^{-8n}\right]\bigg\}.
\end{align}

To analyse the entanglement dynamics in trajectories, we consider two types of unravelings. A naive choice corresponds to a Kraus representation of the channel~\eqref{eq:QC_discrete}:
\begin{align}
    \left.
        \begin{array}{ll}
            &K_0^{(j)} = \sqrt{1-p}~ \mathbb{1}_j,\\
            &K_1^{(j)} = \sqrt{p} \sigma_j^z,
        \end{array}
    \right.
    \label{eq:Kraus_discrete}
\end{align}
to which we refer as $\mathcal{K}^{(j)}$ for brevity. In this case it is easy to see that the middle-cut EAEE~\eqref{eq:EAEE} is linearly growing with $n$ as $\bar{E}(n) = 2n$, since the Kraus operators~\eqref{eq:Kraus_discrete} are both unitary and do not remove any entanglement from the system. 
A better choice for unraveling this dynamics is given by an alternative Kraus representation $\tilde{\mathcal{K}}^{(j)}$ of the channel~\eqref{eq:QC_discrete}:
\begin{align}
    \left.
        \begin{array}{ll}
            &\tilde{K}_0^{(j)} = \ket{0}_j\bra{0} + \left(1-2p\right) \ket{1}_j\bra{1},\\
            &\tilde{K}_1^{(j)} = 2 \sqrt{p\left(1-p\right)}~\ket{1}_j\bra{1},
        \end{array}
    \right.
    \label{eq:Kraus_trans_discrete}
\end{align}
where one of the Kraus operators is a projector onto the excited state, $\tilde{K}_1^{(j)}\propto \ket{1}_j\bra{1}$. Applying such an operator on one qubit removes its entanglement with other qubits. This means that the resulting EAEE is reduced (compared to the EAEE in the naive choice case) and is given by:
\begin{align}
    &\bar{E}_{\tilde{\mathcal{K}}}(p, n) = 2n\left[{q}_{p, n} \log_2 {q}_{p, n} - \tilde{q}_{p, n} \log_2 \tilde{q}_{p, n} + \frac{1}{2} \right], \label{eq:EAEE_analytic}\\
    &\text{with}~q_{p, n} = \frac{1 + (1-2p)^{4n}}{2}~\text{and}~\tilde{q}_{p, n} = \frac{(1-2p)^{4n}}{2}.\nonumber
\end{align}

In Fig.~\ref{fig:4}(b) we plot the evolutions of the entanglement measures for various values of $p$ and for different simulation methods as well as the EoF~\cite{wootters1998entanglement}. From this plot we can draw several conclusions. First, as discussed above, the MPO method as well as the naive trajectory unraveling result in the unbounded growth of the (operator) entanglement, which implies an exponential growth of the bond dimension $\chi$. Second, the alternative unraveling, as well as the 2-GEO adaptive approach, generate trajectories that obey area-law scaling in EAEE. Finally, the 2-GEO unraveling results in the EAEE that is smallest among the considered unraveling schemes and closest to the global minimum $E_f$.

\subsubsection{Continuous case}
\label{sususec:MPO_cont}
The essential features of the above model can also be seen in its continuous-time generalization given by a spin-$1/2$ XX model~\cite{sachdev1999quantum}:
\begin{align}
    H_\mathrm{XX} = \Omega\sum_{j=1}^{L-1} &\left(\sigma_j^+ \sigma_{j+1}^- + \mathrm{h.c.}\right) 
    \label{eq:XX}
\end{align}
whose coupling to Markovian environments is described by the jump operators $c_j = \sigma_j^z / 2$ and decoherence rates $\gamma_j = \gamma$ and where $\sigma_j^\pm = \sigma_j^x \pm i \sigma_j^y$. This model can be solved efficiently numerically after a Jordan-Wigner transformation to a free fermion model~\cite{cao2019entanglement, mbeng2020quantum}.

\begin{figure}[t]\centering
\includegraphics[width=1.\linewidth]{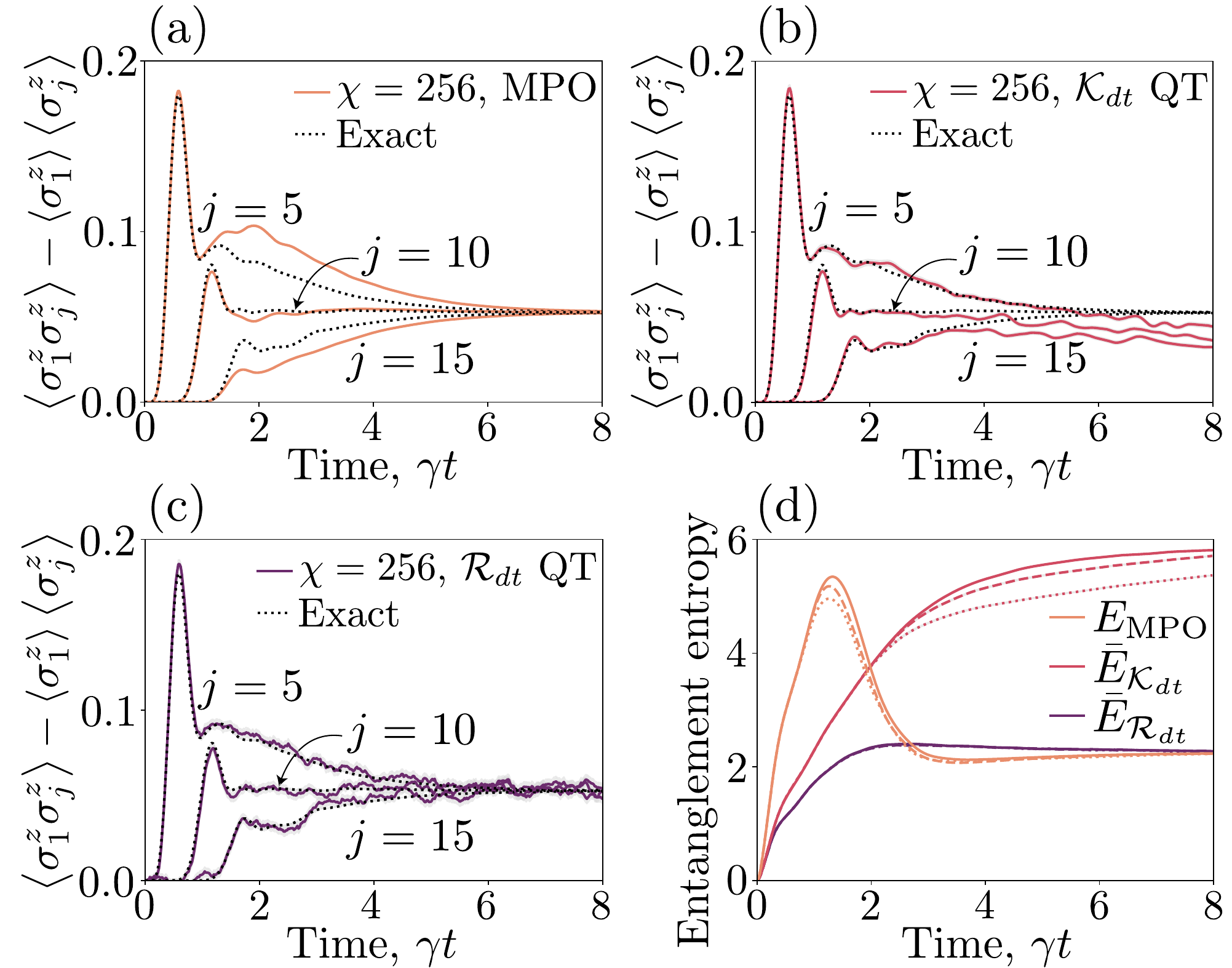}
\caption{Correlation and entanglement dynamics in the XX model~\eqref{eq:XX} for different numerical methods. The system consists of $L=20$ spin-$1/2$ particles and we use jump operators $c_j = \sigma_j^z / 2$ and $\gamma = 0.4 \Omega$. (a), (b) and (c): $\sigma^z$-correlations of the first qubit with the fifth, tenth and fifteenth qubits in the chain simulated with the maximal bond dimension $\chi = 256$ by (a) the MPO method, (b) the ``quantum jump'' QT MPS (total of $M = 2000$ trajectories) and (c) the ``quantum state diffusion'' QT MPS (total of $M = 15000$ trajectories). The dotted black lines correspond to the analytical solution. The statistical error bars in trajectory simulations are denoted by gray filling. Panel (d) compares the dynamics of the (operator) entanglement of the MPO and QT MPS simulations for bond dimensions $\chi = 256$ (solid lines), $\chi = 128$ (dashed lines) and $\chi = 64$ (dotted lines). The EAEE lines of ``quantum state diffusion'' trajectories overlap, which indicates their convergence even for smaller truncation values of the bond dimension.
}
\label{fig:5}
\end{figure}

In Fig.~\ref{fig:5} we solve this model with the MPO and QT MPS methods and compare it to the exact solution\footnote{Due to the dispersion processes in this model the steady-state correlation values are all equal to~$1/L$ and vanish in the thermodynamic limit.}. Specifically, we consider the correlation and entanglement dynamics of $L = 20$ qubits initially in the state~\eqref{eq:state_discrete}. In panels (a), (b) and (c) we fix the truncation bond dimension at $\chi = 256$ and compare the $\sigma^z$-correlation function obtained by the MPO method and the QT MPS unravelings with the ``quantum jump'' Kraus representation $\mathcal{K}_{dt}$ from Eq.~\eqref{eq:Kraus_local} and with the ``quantum state diffusion'' representation $\mathcal{R}_{dt} = T(\theta, \varphi)\mathcal{K}_{dt}$, where $\theta = \pi/4$ and $\varphi = 0$. We can see that for both the MPO and ``quantum jump'' QT MPS approaches the bond dimension is not sufficient to correctly capture the correlation dynamics\footnote{We note that the MPO simulation nevertheless correctly captures the steady state values.}. In contrast, the trajectories obtained by the ``quantum state diffusion'' match the exact results at this value of the bond dimension (within the statistical accuracy). The origin of this difference is visible  in panel (d), where we show the dynamics of the middle-cut (operator) entanglement for various bond dimensions. We observe that the MPO method exhibits an entanglement barrier~\cite{reid2021entanglement,wellnitz2022rise}, a well-known computational issue occurring in quantum many-body system simulations. A similar behaviour can be also seen for the ``quantum jump'' QT MPS method, for which the average entanglement rapidly grows during the evolution and saturates at a value set by the bond dimension. In both cases the lack of convergence in the bond dimension is apparent. In contrast, the average entanglement of ``quantum state diffusion'' trajectories is small and converges with bond dimension.
In the SM~\ref{SM_sec:XX} we consider a more general case of the XX model with fields and give a more detailed discussion on the convergence of observables with the bond dimension.

\section*{Conclusions}
In this work we studied strategies to simulate open quantum many-body system dynamics by combining quantum trajectory methods with tensor-network techniques. The key concept underlying our work is the fact that the same system dynamics can be simulated by infinitely many quantum trajectory unravelings. These unravelings however can significantly differ in the ensemble-averaged entanglement entropy (EAEE) that acts as a proxy of the average computational cost associated with tensor-network representation of individual trajectories. To illustrate this crucial difference, we considered several examples, to which we applied various trajectory unravelings as well as another widely used approach to simulate open quantum many-body dynamics based on matrix product operators (MPOs). These included models that exhibit qualitative differences across different unraveling schemes, as well as models that feature exponential advantage of the trajectory unravelings over MPO simulations. These considerations motivated the development of adaptive trajectory unravelings that minimize the EAEE. We proposed several strategies that utilize this concept by directly minimizing the EAEE (greedy entanglement optimization, GEO) or by minimizing the instantaneous EAEE change rate (greedy entanglement derivative optimization, GEDO, proposed in Ref.~\cite{vovk2022entanglement}). We performed a comprehensive comparison of the GEO and GEDO approaches and concluded that the GEDO method allows to efficiently generate trajectories with low EAEE and that the GEO approach (specifically, 2-GEO) provides trajectory configurations with a lower EAEE value at a cost of less efficient numerical performance. In summary, our work paves the way towards better understanding of the classical simulatability of open quantum many-body systems and defines new avenues for the development of algorithms for noisy quantum many-body dynamics.

There are several directions left for future work. First, our methods are limited to the optimization of individual decoherence channels. It would be interesting to generalize our approach to the collective optimization. Second, it would be interesting to compare or combine the developed methods with other approaches for open system dynamics~\cite{nagy2018driven, nagy2019variational, carisch2023efficient, gravina2023adaptive, muller2024enabling}. Lastly, in the context of measurement-induced phase transitions it would be interesting to determine whether adaptive unravelings of master equations can change the value of the critical point, and, \textit{e.g.}, give area-law trajectories in regimes where non-adaptive unravelings can not.

~
~
~

\section*{Acknowledgements}
We thank Lukas Sieberer for carefully reading the manuscript draft and giving constructive feedback. We thank Peter Zoller, Barbara Kraus, Tom Westerhout, Matteo Magoni, Luca Arceci, and Gabriele Calliari for fruitful discussions. The computational results presented here have been achieved using the LinuX Cluster of the Institute for Theoretical Physics of the University of Innsbruck and LEO HPC infrastructure of the University of Innsbruck. This work is supported by the European Union’s Horizon Europe research and innovation program under Grant Agreement No. 101113690 (PASQuanS2.1), the ERC Starting grant QARA (Grant No.~101041435), the EU-QUANTERA project TNiSQ (N-6001), and by the Austrian Science Fund (FWF): COE 1 and quantA. For open access purposes, the author has applied a CC BY public copyright license to any author accepted manuscript version arising from this submission.

\bibliography{bibliography} 

\clearpage
\pagebreak
\widetext
\begin{center}
\textbf{\large Supplementary Material}
\end{center}
\setcounter{equation}{0}
\setcounter{figure}{0}
\setcounter{table}{0}
\setcounter{section}{0}
\setcounter{footnote}{0}
\makeatletter
\renewcommand{\theequation}{S\arabic{equation}}
\renewcommand{\thefigure}{S\arabic{figure}}
\renewcommand{\bibnumfmt}[1]{[S#1]}

\section{Optimization over $T\in\mathcal{T}_2^2$}
\label{SM_sec:T22}
In this section we provide the details of the numerical solution of the optimization problem~\eqref{eq:E_new}. Here we consider a limited class of isometries $T(\theta, \varphi)\in\mathcal{T}_2^2\subset \mathcal{T}_2^4$ explicitly given in~\eqref{eq:2x2_T}. Applying such an isometry to the reference ensemble $\tilde\Psi_{dt}= \mathcal{K}_{dt} \ket{\psi} = \left(K_0\ket{\psi}, K_1\ket{\psi}\right)$ results in a new ensemble of the form\footnote{For simplicity in what follows we assume the Kraus operators and corresponding jump operators to be local.}:
\begin{align}
    \begin{pmatrix}
        \ket{\tilde{\phi}_0\left(\theta, \varphi, dt\right)} \\
        \ket{\tilde{\phi}_1\left(\theta, \varphi, dt\right)} \\
    \end{pmatrix} = 
    T(\theta, \varphi)
    \begin{pmatrix}
        K_0\ket{\psi} \\
        K_1\ket{\psi} \\
    \end{pmatrix}.
    \label{eq:2x2}
\end{align}
Inserting it into the EAEE~\eqref{eq:EAEE} allows to reformulate the optimization problem~\eqref{eq:E_new} as
\begin{align}
    \inf_{T\in\mathcal{T}_2^4} \bar{E}[T \tilde{\Psi}_{dt}] \xrightarrow{T=T(\theta, \varphi)} \min_{\theta, \varphi} \bar{E}\left(\theta, \varphi, dt\right),
    \label{eq:E_new_parametrised}
\end{align}
the solution of which we aim to explain below. The EAEE $\bar{E}\left(\theta, \varphi, dt\right)$ in Eq.~\eqref{eq:E_new_parametrised} is convex, which allows to define unambiguously the minimum using the standard optimization methods. We choose the Broyden-Fletcher-Goldfarb-Shanno (BFGS) algorithm -- a quasi-Newton gradient-based method, which efficiently solves nonlinear optimization problems~\cite{broyden1970convergence,fletcher1970new,goldfarb1970family,shanno1970conditioning}. Below we provide an explicit form of the EAEE as a function of the optimization parameters and contraction schemes of MPS networks needed for BFGS optimization.

Let us write the EAEE as a function of the isometry parameters\footnote{From here on we do not write $dt$ for notational clarity.}:
\begin{align}
    \bar{E}\left(\theta, \varphi\right) = \sum_{\alpha=0,1} p_\alpha\left(\theta, \varphi\right) E\left(\ket{{\phi}_\alpha\left(\theta, \varphi\right)}\right),
\end{align}
where $p_\alpha\left(\theta, \varphi\right) = \braket{\tilde{\phi}_\alpha\left(\theta, \varphi\right)}{\tilde{\phi}_\alpha\left(\theta, \varphi\right)}$ are the probability weights of the ensemble~\eqref{eq:2x2} and $E\left(\ket{{\phi}_\alpha\left(\theta, \varphi\right)}\right)$ is the von Neumann entanglement entropy~\eqref{eq:vN} of the normalized state $\ket{{\phi}_\alpha\left(\theta, \varphi\right)} = \ket{\tilde{\phi}_\alpha\left(\theta, \varphi\right)} / \sqrt{p_\alpha\left(\theta, \varphi\right)}$. 
If we define the partial trace $\tilde{\phi}_{\alpha} \left(\theta, \varphi\right) = \mathrm{tr}_B \left(\ket{\tilde{\phi}_\alpha\left(\theta, \varphi\right)}\bra{\tilde{\phi}_\alpha\left(\theta, \varphi\right)}\right)$, we can write:
\begin{align}
    \bar{E}\left(\theta, \varphi\right) = \sum_{\alpha=0,1} p_\alpha\left(\theta, \varphi\right) \log_2 p_\alpha\left(\theta, \varphi\right) - \mathrm{tr}\left[\tilde{\phi}_{\alpha}\left(\theta, \varphi\right) \log_2\left(\tilde{\phi}_{\alpha}\left(\theta, \varphi\right)\right)\right].
    \label{eq:E}
\end{align}
The gradients of such a function have a simple form:
\begin{align}
    \partial_{\theta, \varphi}\bar{E}\left(\theta, \varphi\right) = - \sum_{\alpha=0,1} \mathrm{tr}\left[\partial_{\theta, \varphi}\tilde{\phi}_{\alpha}\left(\theta, \varphi\right) \log_2\left({\phi}_{\alpha}\left(\theta, \varphi\right)\right)\right],
    \label{eq:E_grad}
\end{align}
where ${\phi}_{\alpha}\left(\theta, \varphi\right) = \tilde{\phi}_{\alpha}\left(\theta, \varphi\right)/p_\alpha\left(\theta, \varphi\right)$. One can find the explicit forms of the probability weights $p_\alpha\left(\theta, \varphi\right)$ and the partial traces $\tilde{\phi}_{\alpha}\left(\theta, \varphi\right)$ in terms of the optimization parameters $\theta, \varphi$ and the initial ensemble $\tilde\Psi_{dt} = (K_0\ket{\psi}, K_1\ket{\psi}) = (\ket{\tilde\psi_0}, \ket{\tilde\psi_1})$:
\begin{align}
    &p_0\left(\theta, \varphi\right) = \cos^2(\theta) q_0 + \sin^2(\theta) q_1 + 2 \cos(\theta)\sin(\theta)\cos(\varphi - \upsilon_{01}) q_{01},\\
    &p_1\left(\theta, \varphi\right) = \sin^2(\theta) q_0 + \cos^2(\theta) q_1 - 2 \cos(\theta)\sin(\theta)\cos(\varphi - \upsilon_{01}) q_{01},\\
    &\tilde{\phi}_0\left(\theta, \varphi\right) = \cos^2(\theta) \tilde{\psi}_0 + \sin^2(\theta) \tilde{\psi}_1 + \cos(\theta)\sin(\theta) \left(e^{-i\varphi}\tilde{\psi}_{10} + \mathrm{h.c.}\right),\\
    &\tilde{\phi}_1\left(\theta, \varphi\right) = \sin^2(\theta) \tilde{\psi}_0 + \cos^2(\theta) \tilde{\psi}_1 - \cos(\theta)\sin(\theta)\left(e^{-i\varphi}\tilde{\psi}_{10} + \mathrm{h.c.}\right).
\end{align}
where $q_\alpha = \braket{\tilde{\psi}_{\alpha}}{\tilde{\psi}_{\alpha}}$ are the probability weights of the initial ensemble, $q_{01} = |\braket{\tilde{\psi}_0}{\tilde{\psi}_1}|$ and $\upsilon_{01} = \arg\left(\braket{\tilde{\psi}_0}{\tilde{\psi}_1}\right)$ are amplitude and phase of the overlap of the initial ensemble states,  $\tilde{\psi}_\alpha = \mathrm{tr}_B \left(\ketbra{\tilde{\psi}_{\alpha}}{\tilde{\psi}_{\alpha}}\right)$ are the partial traces of the initial ensemble and $\tilde{\psi}_{10} = \mathrm{tr}_B \left(\ketbra{\tilde{\psi}_1}{\tilde{\psi}_0}\right)$. With this one can calculate the derivatives of the matrices:
\begin{align}
    &\partial_\theta \tilde{\phi}_0\left(\theta, \varphi\right) = 2 \cos(\theta) \sin(\theta) \left(\tilde{\psi}_1 - \tilde{\psi}_0\right) + 2 \left[\cos^2(\theta) - \sin^2(\theta)\right] \left(e^{-i\varphi}\tilde{\psi}_{10} + \mathrm{h.c.}\right),\\
    &\partial_\theta \tilde{\phi}_1\left(\theta, \varphi\right) = 2 \cos(\theta) \sin(\theta) \left(\tilde{\psi}_0 - \tilde{\psi}_1\right) - 2 \left[\cos^2(\theta) - \sin^2(\theta)\right] \left(e^{-i\varphi}\tilde{\psi}_{10} + \mathrm{h.c.}\right), \\
    &\partial_\varphi \tilde{\phi}_0\left(\theta, \varphi\right) = i \cos(\theta) \sin(\theta) \left(- e^{-i\varphi}\tilde{\psi}_{10} + e^{i\varphi}\tilde{\psi}_{01}\right),\\
    &\partial_\varphi \tilde{\phi}_1\left(\theta, \varphi\right) = i \cos(\theta) \sin(\theta) \left(e^{-i\varphi}\tilde{\psi}_{10} - e^{i\varphi}\tilde{\psi}_{01}\right),
\end{align}
which can be inserted into~\eqref{eq:E_grad} to calculate the gradient.

\begin{figure}[h!]\centering
\includegraphics[width=0.85\linewidth]{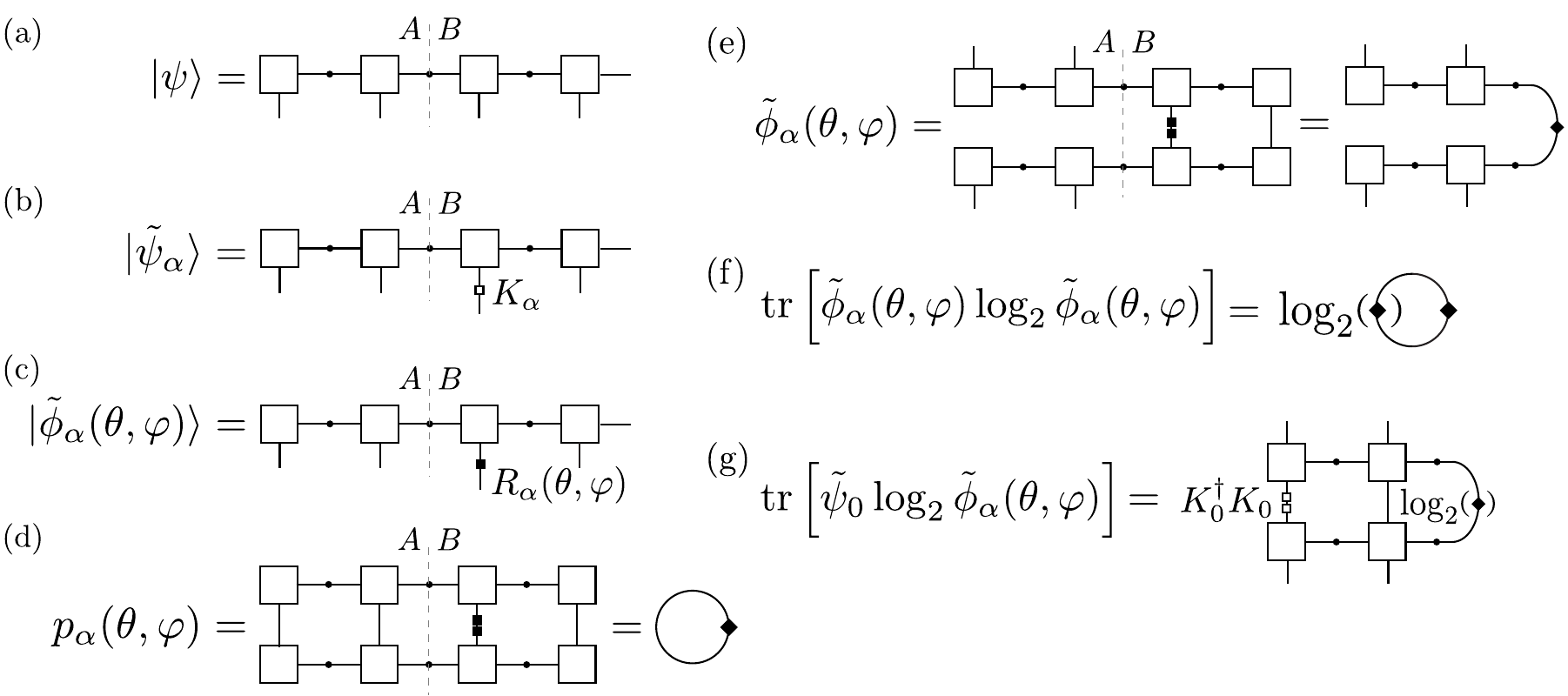}
\caption{The MPS tensor objects in Vidal's $\Gamma \lambda$-form~\cite{vidalEfficientClassicalSimulation2003} and their contractions required to calculate the EAEE and it gradients.
(a) The initial state $\ket{\psi}$ with the bipartition $AB$. (b) The initial ensemble state $\ket{\tilde{\psi}_{\alpha}}$ obtained by applying the corresponding Kraus operator $K_\alpha$. (c) The transformed ensemble state $\ket{\tilde{\phi}_{\alpha}\left(\theta, \varphi\right)}$ with $R_\alpha (\theta, \varphi) = \sum_\beta T_{\alpha\beta}(\theta, \varphi) K_\beta$. (d) The probability weight $p_\alpha\left(\theta, \varphi\right)$. (e) The partial trace $\tilde{\phi}_{\alpha} \left(\theta, \varphi\right) = \mathrm{tr}_B \left(\ket{\tilde{\phi}^{(\alpha)}\left(\theta, \varphi\right)}\bra{\tilde{\phi}^{(\alpha)}\left(\theta, \varphi\right)}\right)$. (f) The term $\mathrm{tr}\left[\tilde{\phi}_{\alpha}\left(\theta, \varphi\right) \log_2\left(\tilde{\phi}_{\alpha}\left(\theta, \varphi\right)\right)\right]$ from~\eqref{eq:Schmidt_trick}. (g) The term $\mathrm{tr}\left[\tilde{\psi}_0 \log_2\left({\phi}_{\alpha}\left(\theta, \varphi\right)\right)\right]$ as an example of a building block to calculate the EAEE gradient~\eqref{eq:E_grad}.
}
\label{fig:SM1}
\end{figure}

In the case when the system state is represented in the MPS form, one can calculate the EAEE~\eqref{eq:E} and its gradients~\eqref{eq:E_grad} by contracting tensor networks as shown in Fig.~\ref{fig:SM1}. There are two important aspects to notice. First, this optimization is numerically costly when using MPS contractions. The reason for this is that for every set of parameters $\theta, \varphi$ one has to calculate singular values of the MPS for every new optimization iteration, since these values depend implicitly on the optimizing parameters. This can be seen by analyzing the second term of Eq.~\eqref{eq:E}. This term can be written using the singular value decomposition as:
\begin{align}
    \mathrm{tr}\left[\tilde{\phi}_{\alpha}\left(\theta, \varphi\right) \log_2\left(\tilde{\phi}_{\alpha}\left(\theta, \varphi\right)\right)\right] = \sum_i \tilde{\lambda}^{(i)}_\alpha \left(\theta, \varphi\right) \log_2 \tilde{\lambda}^{(i)}_\alpha \left(\theta, \varphi\right),
    \label{eq:Schmidt_trick}
\end{align}
where $\tilde{\lambda}^{(i)}_\alpha \left(\theta, \varphi\right)$ are the eigenvalues of $\tilde{\phi}_{\alpha} \left(\theta, \varphi\right)$. Second, in the case of the gradient calculation~\eqref{eq:E_grad}, in addition to singular value calculation one has to explicitly contract the corresponding tensor networks instead of just using the singular values [see Fig.~\ref{fig:SM1}(g)].

\section{Comment on optimization over $T\in\mathcal{T}_2^4$}
\label{SM_sec:T24}
So far we covered the EAEE optimization over a limited class of isometries $T\in\mathcal{T}_2^2$. However, in order to guaranteedly attain the minimum~\eqref{eq:E_new}, one has to consider the full class of isometries $\mathcal{T}_2^4$~\cite{hedemann2013hyperspherical}. This implies optimizations of $3\times2$ right-unitary matrices with five parameters:
\begin{align}
    T(\theta_1, \theta_2, \theta_3, \alpha, \beta) = 
    \begin{pmatrix}
        \cos(\theta_1) && \sin(\theta_1) \cos(\theta_2) e^{i\alpha}\\
        \sin(\theta_1) \cos(\theta_3) & \begin{split}
         &- \cos(\theta_1) \cos(\theta_2) \cos(\theta_3) e^{i\alpha}
         \\
       &- \sin(\theta_2) \sin(\theta_3) e^{-i\beta} 
        \end{split} \\
        \sin(\theta_1) \sin(\theta_3) & \begin{split}
         &- \cos(\theta_1) \cos(\theta_2) \sin(\theta_3) e^{i\alpha}
         \\
        &+ \sin(\theta_2) \cos(\theta_3) e^{-i\beta}
        \end{split} \\
    \end{pmatrix},
    \label{eq:3x2_T}
\end{align}
and of $4\times2$ right-unitary matrices with eight parameters:
\begin{align}
    &T(\theta_1, \theta_2, \theta_3, \theta_4, \theta_5, \alpha, \beta, \gamma) =\nonumber\\
    &~~~~\begin{pmatrix}
        \cos(\theta_1) && \sin(\theta_1) \cos(\theta_2) e^{i\alpha}\\
        \sin(\theta_1) \cos(\theta_3) && \sin(\theta_2) \sin(\theta_3) \cos(\theta_5) e^{i\beta} - \cos(\theta_1) \cos(\theta_2) \cos(\theta_3) e^{i\alpha} \\
        \sin(\theta_1) \sin(\theta_3) \cos(\theta_4) & \begin{split}
         &\sin(\theta_2) \sin(\theta_4) \sin(\theta_5) e^{i\gamma}  - \cos(\theta_1) \cos(\theta_2) \sin(\theta_3) \cos(\theta_4) e^{i\alpha}\\
         & ~~~~~~- \sin(\theta_2) \cos(\theta_3) \cos(\theta_4) \cos(\theta_5) e^{i\beta}
        \end{split} \\
        \sin(\theta_1) \sin(\theta_3) \sin(\theta_4) &\begin{split}
         &~~~-\cos(\theta_1) \cos(\theta_2) \sin(\theta_3) \sin(\theta_4) e^{i\alpha} - \sin(\theta_2) \cos(\theta_3) \sin(\theta_4) \cos(\theta_5) e^{i\beta} \\
         & ~~~~~~~~~~~-\sin(\theta_2) \cos(\theta_4) \sin(\theta_5) e^{i\gamma}
        \end{split} \\
    \end{pmatrix}.
    \label{eq:4x2_T}
\end{align}
The EAEE and its gradients associated with the isometries~\eqref{eq:3x2_T} and ~\eqref{eq:4x2_T} have a more complicated form than the ones shown in Section~\ref{SM_sec:T22}, but the calculations are similar. The important aspect to notice in this case is that the high-dimensional optimization landscapes for~\eqref{eq:3x2_T} and~\eqref{eq:4x2_T} make the optimization process more complicated.

\section{Numerical analysis of single-step optimization: A case of ten qubits}
\label{SM_sec:10qubit}

\begin{figure}[b]\centering
\includegraphics[width=0.4\linewidth]{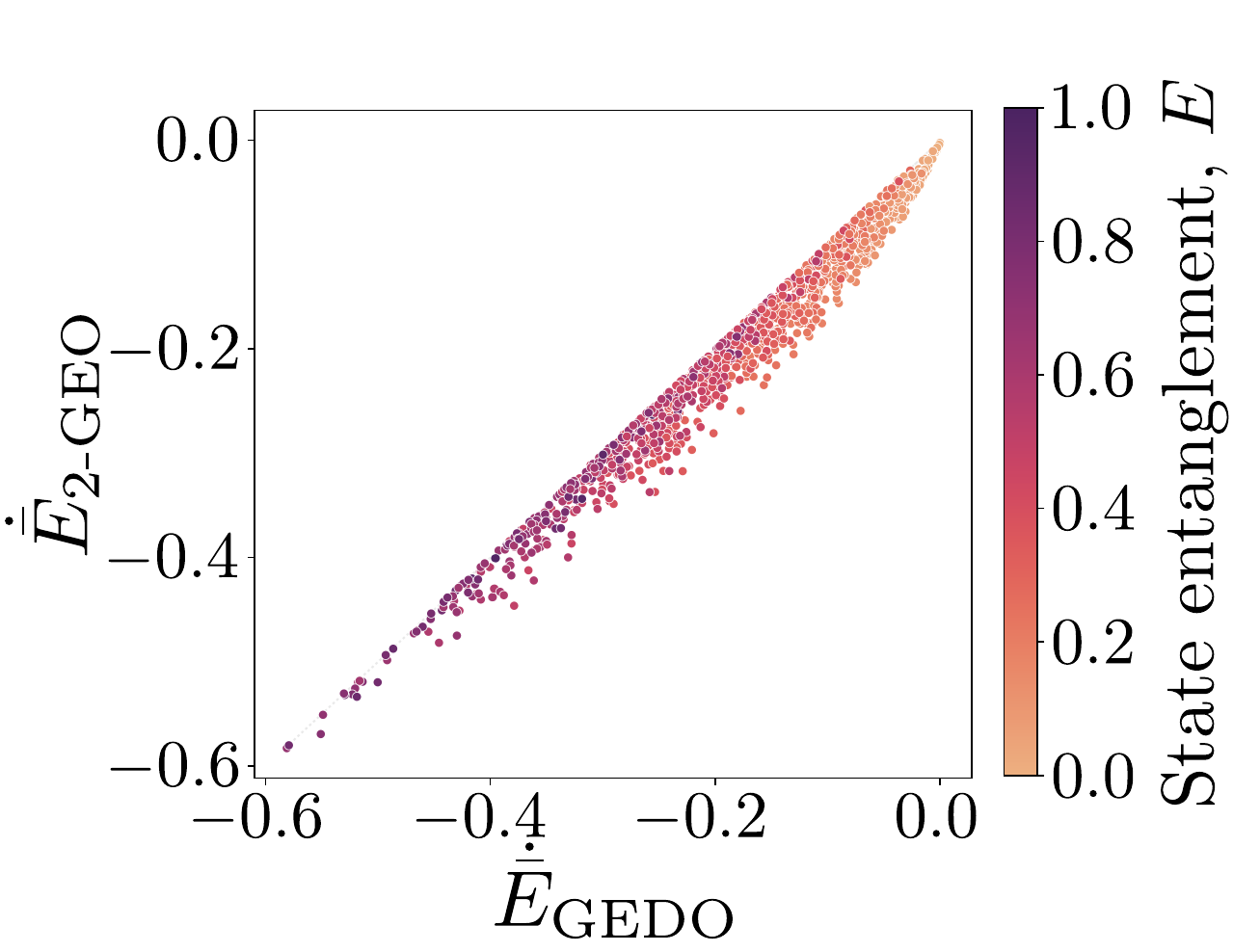}
\caption{Middle-cut EAEE change rates of the system of $L=10$ qubits obtained by the 2-GEO vs. GEDO methods with a single jump operator $c_j = \sigma_j^z / 2$, $\gamma_j = 1$ and $j = L/2$. Each dot represents a random state (total of 1250 instances).
}
\label{fig:SM2}
\end{figure}

In Subsubsection~\ref{susec:comparison} of the main text we compare the performance of the adaptive unraveling strategies, namely, the 2-GEO and GEDO methods (see Subsection~\ref{susec:adaptive} of the main text). Here we extend our analysis to larger systems. For this we consider a system of $L=10$ qubits and perform the same procedure as in Fig.~\ref{fig:1}(b). Namely, we take typical states of the 10-qubit system, let them evolve under a decoherence channel for a single time step and compare the resulting EAEE change rates for different adaptive unravelings. Note that, in contrast to the two-qubit scenario discussed in the main text, the 10-qubit system can be partitioned in multiple ways, and, hence, many bipartite entanglement values may be optimized. Here we focus on the case of the middle-cut bipartition and corresponding EAEE values.

As typical states we take a class of random states with various middle-cut entanglement entropies. These states are generated by propagating a product state of excited qubits $\bigotimes_{i=1}^L\ket{1}_i$ with a time-dependent random Hamiltonian defined in Eq.~\eqref{eq:RBC}. This fast-scrambling Hamiltonian increases the entanglement in the qubit system linearly in time. Therefore, one can explore both weakly and strongly entangled many-body states by changing the propagation times. In what follows we take 25 different propagation times starting from $t_\mathrm{min} = 0.2 / \gamma_j$ and up to $t_\mathrm{max} = 5 / \gamma_j$, where $\gamma_j = 1$. The propagation time step is $dt = 0.01 / \gamma_j$. For each of the propagation durations we apply 50 different random Hamiltonian realizations, which in total gives us 1250 random many-body states with different entanglement entropies.

In Fig.~\ref{fig:SM2} we compare the EAEE values of the 2-GEO and GEDO unravelings of the 10-qubit evolution under the decoherence channel with $c_{j} = \sigma_j^z / 2$, where $j = 5$. From this plot we draw a conclusion similar to the two-qubit case: the 2-GEO method finds configurations that give lower EAEE values compared to the GEDO method. We note that, in contrast to the two-qubit case, in the 10-qubit case the entanglement change rate is no longer monotonically related to the value of initial-state entanglement entropy, which can be seen in the distribution of colors in Fig.~\ref{fig:SM2}.

\section{XX model with longitudinal and transverse fields}
\label{SM_sec:XX}
\begin{figure}[b]\centering
\includegraphics[width=1\linewidth]{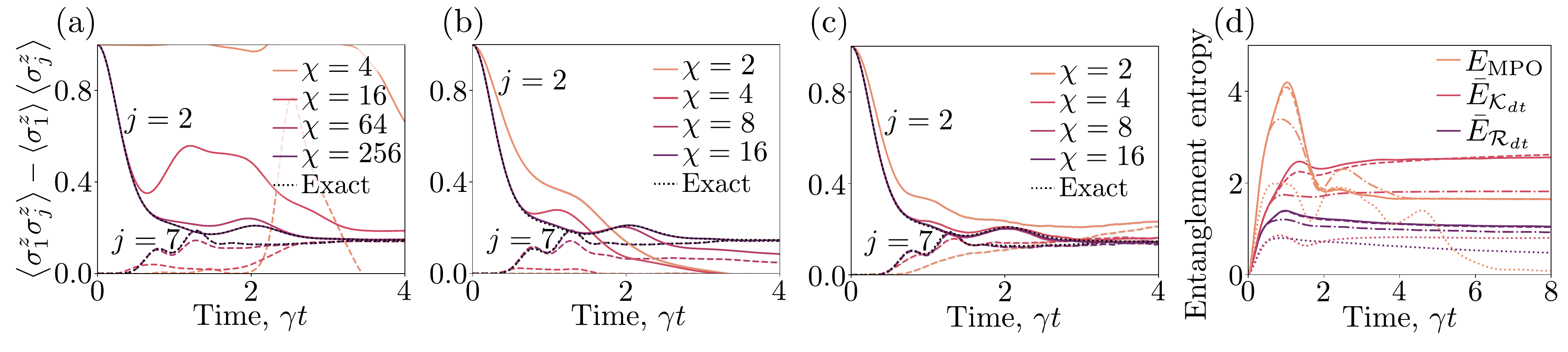}
\caption{Correlation and entanglement dynamics in the noisy XX model with fields~\eqref{eq:XXF} for different numerical methods. The system consists of $L=8$ spin-$1/2$ particles and we use jump operators $c_j = \sigma_j^z / 2$ and $\gamma = 0.4 \Omega$, $\Delta = h = 0.02 \Omega$. (a), (b) and (c): $\sigma^z$-correlations of the first qubit with the second qubit (solid lines) and with the seventh qubit (dashed lines) simulated with various bond dimensions $\chi$ by (a) the MPO method, (b) the ``quantum jump'' QT MPS (total of $M = 10000$ trajectories) and (c) the ``quantum state diffusion'' QT MPS (total of $M = 20000$ trajectories). The dotted black lines correspond to the exact solution. In this case the error bars are not exceeding the lines' width and hence are not shown. Panel (d) compares the dynamics of the (operator) entanglement of the MPO method, shown in bright orange color with $\chi = 256$ (solid line), $64$ (dashed line), $16$ (dash-dotted line) and $4$ (dotted line), and the QT MPS methods, shown in pink and dark purple colors with $\chi = 16$ (solid lines), $8$ (dashed lines), $4$ (dash-dotted lines) and $2$ (dotted lines).
}
\label{fig:SM3}
\end{figure}

Complementing the study in Subsubsection~\ref{sususec:MPO_cont} of the main text, here we consider the noisy XX model with fields:
\begin{align}
    H_\mathrm{XXF} =  \Omega\sum_{j=1}^{L-1} &\left(\sigma_j^+ \sigma_{j+1}^- + \mathrm{h.c.}\right) + \Delta \sum_{j=1}^{L}  \sigma_j^z+ h \sum_{j=1}^{L} \sigma_j^x,
    \label{eq:XXF}
\end{align}
which cannot be mapped to free fermions and therefore is not efficiently numerically solvable. In Fig.~\ref{fig:SM3} we solve this model with the MPO and QT MPS methods and compare it to the exact diagonalization solution. Specifically, we consider the correlation and entanglement dynamics in a system of $L = 8$ qubits initially in the state~\eqref{eq:state_discrete}. In panels (a), (b) and (c) we plot the $\sigma^z$-correlations obtained by the MPO method and the QT MPS ``quantum jump'' and ``quantum state diffusion'' unravelings. In each panel we vary the value of the maximal bond dimension $\chi$ and check the convergence of the simulations towards the exact diagonalization solution. We set the maximal MPO bond dimension to be the square of the analogous value for the MPSs for comparison clarity~\cite{van2017dynamics}.

We observe that for a fixed maximal bond dimension $\chi = 16$ the QT MPS simulations give exact results, whilst the MPO method significantly deviates from the exact solution even for a larger maximal bond dimension $\chi = 64$. In turn, comparing the two trajectory approaches, we see that ``quantum jump'' trajectories also significantly deviate from the correct results once the maximal bond dimension is reduced. In contrast to that, the ``quantum state diffusion'' trajectories are closer to the exact results for lower $\chi$ values. The origin of these differences is visible in panel (d), where we plot the dynamics of the middle-cut (operator) entanglement for various bond dimensions. We observe that for a fixed bond dimension $\chi = 16$ the MPO method is not exact in contrast to the trajectory methods. In addition, reducing the maximal bond dimension to $\chi = 8$ makes ``quantum jump'' trajectories deviate from the exact line, which does not take place for ``quantum state diffusion'' trajectories.

\end{document}